\documentclass[pra,superscriptaddress,twocolumn,nofootinbib]{revtex4-2}

\usepackage{graphicx,epsfig}
\usepackage{color}
\usepackage[usenames,dvipsnames]{xcolor}
\usepackage{amsmath,bbm,amssymb, amsthm}
\usepackage{dsfont} 
\usepackage{stmaryrd}
\usepackage{comment}
\usepackage{float}
\usepackage{braket}
\usepackage{hyperref}
\usepackage{numprint}

\def\bb{\begin{eqnarray}}
\def\ee{\end{eqnarray}}
\renewcommand{\ket}[1]{| #1 \rangle}
\renewcommand{\bra}[1]{\langle #1 |}

\newcommand{\moy}[1]{\left\langle #1 \right\rangle} 

\newcommand{\idop}{\mathds{1}}

\newcommand{\red}[1]{{\color[rgb]{0,0,0}{#1}}}
\newcommand{\Cyril}[1]{{\color[rgb]{0,0,1}{#1}}}

\begin{document}
\title{The Jaynes-Cummings model as an autonomous Maxwell demon}

\author{Yashovardhan Jha}
\email{yashoovardhanjha@gmail.com} 
\affiliation{Université de Lorraine, CNRS, LPCT, F-54000 Nancy, France}

\author{Dragi Karevski}
\affiliation{Université de Lorraine, CNRS, LPCT, F-54000 Nancy, France}

\author{Cyril Elouard}
\email{cyril.elouard@univ-lorraine.fr} 
\affiliation{Université de Lorraine, CNRS, LPCT, F-54000 Nancy, France}

\begin{abstract}
We revisit the Jaynes-Cummings model as an autonomous thermodynamic machine, where a qubit is driven by a cavity containing initially a large coherent field. Our analysis reveals a transition between the expected behavior of ideal-work source of the cavity at short times, and a long-time dynamics where the cavity autonomously measures the qubit and exerts a result-dependent drive. This autonomous feedback then purifies the qubit irrespective of its initial state. We show that the cavity functions thermodynamically as an autonomous Maxwell demon, trading mutual information for cooling power.
\end{abstract}

\maketitle

\section{Introduction}


\red{Autonomous machines are} a key concept in thermodynamics \cite{Toyabe20,Guzman24}. \red{Their time evolution solely derives from their initial non-equilibrium condition, akin to an automaton,} allowing them to perform tasks without external control. This concept is all the more important in the case of quantum machines \cite{Tonner05,Brask15,Roulet17,Monsel18,Latune19,Manzano19,Rasola24} where the coherence of microscopic constituents is crucial, \red{as it avoids employing classical controllers which are typically sources of decoherence.}
The latter observation motivates research into autonomous (or coherent) feedback schemes \cite{Wiseman94,Nelson00,Lloyd00,Rouchon14,Zhou15,Wolf23}, that is, protocols where a quantum system is able to drive a target system conditionally on the values of a few of its observables, e.g. applied to quantum error correction \cite{Reiter17,Xu23,Lachance-Quirion24}. 
When these feedback controllers perform thermodynamic tasks, such as work extraction or cooling, they are called autonomous Maxwell demons \cite{Barato13,Strasberg18,Najera-Santos20,Ciliberto20,Freitas21,Annby-Andersson24,Monsel25}. \red{Like in the famous gedanken experiment attributed to Maxwell, the information about the system exploited in the feedback mechanism is a resource, enabling work extraction from a single heat source or cold-to-hot heat transfer in the absence of work, hence apparent violations of the second law if those resources are disregarded.}
A thorough characterization of these machines is of major fundamental interest, as it is linked to the long-standing question of how the dynamics induced by quantum measurements \cite{Allahverdyan13,Engineer24,Latune25} and the laws of thermodynamics \cite{Strasberg21,Elouard23,Meier25} emerge in an entirely quantum framework.


Here, we analyze an autonomous machine composed of a qubit interacting with a cavity initialized in a coherently displaced thermal state, focusing on the classical limit. Over time, we identify three successive regimes exhibiting very different thermodynamic behaviors. At short times, the cavity behaves as a quasi-ideal source of work, namely, a classical drive inducing unitary Rabi oscillations of the qubit. We quantify the deviations from ideality, which take the form of residual heat exchange and entropy production. This regime breaks down over a time-scale set by the qubit-cavity coupling constant, where Rabi oscillations decohere. This decoherence is related to an autonomous measurement of the qubit by the cavity in a basis set by the initial phase of the field. The result of this measurement is stored in a conditional field amplitude. Finally, we identify a third regime where the cavity unitarily drives the qubit depending on these measurement outcomes. This autonomous feedback is able to purify the qubit starting from any mixed initial state. Our thermodynamic analysis shows that the cavity behaves as an autonomous Maxwell demon, trading mutual information for cooling power.

\section{Thermodynamics of autonomous quantum machines}\label{s:Autothermo}
To interpret the energy exchanges between the cavity and the qubit in terms of work and heat, and deduce the constraints set by the second law of thermodynamics, we rely on the framework introduced by some of us in Ref.~\cite{Elouard23}. \red{In a nutshell, this framework identifies work with energy exchanges along isoentropic transformation, and heat with energy exchanges proportional to the entropy variations}. More precisely, the method applies to an arbitrary set of quantum systems whose evolution is ruled by a total time-independent Hamiltonian
\bb
 \hat H = \sum_j \hat H_j + \hat V ,
\ee
where $H_j$ stands for the bare Hamiltonian of system $j$, and $\hat V$ an arbitrary interaction Hamiltonian. Those systems can include the thermodynamic system of interest (the working body), as well as the energy sources used to power its transformation, \red{treated as quantum systems}. Importantly, the role of each system $j$ (e.g. heat source or work source) is not pre-supposed; all systems are a priori hybrid sources of work and heat for the others, and are treated on an equal footing.

The internal energy of system $j$ is defined as $E_j(t)=\text{Tr}\{\hat H_j\hat\rho_j(t)\}$, computed from its reduced density operator $\hat\rho_{j}(t)= \text{Tr}_{j'\neq j}\{\hat\rho_\text{tot}(t)\}$, where $\hat\rho_\text{tot}(t)$ stands for the density \red{operator} of the whole set of systems. Then, the heat provided by system $j$ to the others between times $t=0$ and $t$ is defined as the variation 
\bb
 Q_j(t)=-\Delta E_j^\text{th} ,
\ee
of the thermal energy
\bb
 E_j^\text{th}(t) = \text{Tr}\{\hat w_j(t) \hat H_j\}.
\ee
The latter is the internal energy of the \red{unique} thermal state $\hat w_j(t)=e^{-\beta_j(t)\hat H_j}/Z_j(t)$ of system $j$ which has the same von Neumann entropy 
\bb
S_j(t)=-
\text{Tr}\{\hat\rho_j(t)\log\hat\rho_j(t)\}
\ee
as the actual reduced state $\hat\rho_j(t)$. As thermal states achieve minimal energy at fixed von Neumann entropy, the thermal energy $E_j^\text{th}(t)\leq E_j(t)$ represents the part of the internal energy of system $j$ which carries entropy (or the energy remaining in the system after \red{performing on it} the isoentropic transformation extracting the largest amount of energy). For an infinitesimal evolution, \red{the heat} verifies
\bb
\red{\delta} Q_j(t) = dt \dot{Q}_j(t)=-\beta_j^{-1}(t)dS_j(t),
\ee
such that an exchange of heat is necessarily accompanied by a change of entropy of the energy source providing heat. 
Conversely, the work provided by the system $j$ is \red{defined as the} energy exchange associated to the isoentropic part of the evolution of system $j$ and is obtained from:
\bb\label{workj}
W_j(t) = \red{-}\Delta E_j-Q_j(t).
\ee
\red{The effective nonequilibrium inverse temperature $\beta_j(t)$ is the solution of the equation $S_j(t)=-\text{Tr}\{\hat w_j(t)\log \hat w_j(t)\}$. It sets constraints on the energy exchanges occurring during the joint evolution of the systems}. For instance, for an uncorrelated total initial state of the systems $\hat\rho_\text{tot}(\red{t_0})=\otimes_j \hat\rho_j(\red{t_0})$, one can express the second law for the transformation of system $j$ under the form \red{of Clausius inequality} \cite{Elouard23}
\red{
\bb
 \Delta S_j - \sum_{j'\neq j} \beta_{j'}(0) {Q}_{j'}(t) \geq 0. \label{entprodj}
\ee }
From these definitions, a pure source of work, verifying $\Delta E_j = W_j(t)$, can be identified as a quantum system never building correlations with the others, that is, fulfilling \red{$\hat\rho_\text{tot}(t)= \hat\rho_j(t)\otimes \text{Tr}_j\{\hat\rho_\text{tot}(t)\}$}. The evolution of system $j$ is then isoentropic, and it effectively behaves as a classical drive on the remaining systems. Such behavior is expected \red{to occur only asymptotically in realistic systems, such as a cavity in the classical limit $n_0 \gg 1, \bar n$, coupled to an atom. The framework presented here quantitatively evaluates deviations from this ideal picture, e.g. residual exchanges of heat associated with the finite perturbation of the cavity state and the correlations generated by the cavity-atom dynamics}.\\

\section{(Autonomous) Maxwell demons}\label{s:MaxwellDemons}

\red{In a modern formulation \cite{Horowitz14,Sagawa11,Parrondo23,Sanchez19}, Maxwell demons can be defined as entities acquiring information on the quantum system of interest and performing a feedback on it based on this information. If the presence of the demon is not known, and the system only is monitored, the action of the demon results in apparent violations of the second law, such as a decrease of the system's entropy without an outward heat flow or a cyclic work extraction from a single heat source.

In the historical (non-autonomous) scenario, i.e. Maxwell's seminal gedanken experiment, the demon is an agent performing a measurement on the system and then storing the result in a memory. It then acts conditionally to this result on the system, allowing for an apparent cyclic work extraction from a single heat source. The paradox is solved by considering the finite work cost to reset the demon's memory, which exactly compensates the extracted work when the reset process is performed in contact with the same heat source as the work extraction. Alternatively, in non-cyclic transformations, one must consider the information previously acquired by the demon as a resource, which when consumed, plays a role similar to a work expenditure.
A similar analysis can be done when the system and the demon are described together as an autonomous system, solely evolving due to the dynamics generated by their joint, time-independent Hamiltonian: in this case, information flows from the system to the demon (which can be seen as the autonomous counterpart of a measurement) and influences the later dynamics induced by the demon on the working system (autonomous feedback). 
The dynamics of the system under the influence of the demon obeys a modified Clausius inequality, taking into account the information acquired by the demon prior to time $t_0$. Focusing on a system S interacting with a demon D, both forming an isolated system, one has \cite{Elouard23}:
\bb \label{entproddemon}
\Delta S_S - \beta_D(0)Q_D(t) \geq \Delta I(t).
\ee 
Here $I(t) = S_S(t)+S_D(t)-S_{SD}(t) \geq 0$ is the quantum mutual information between the demon and the system at time $t$, and $\Delta I(t)=I(t)-I(t_0)$ its variation between time $t_0$ and $t>t_0$. $I(t_0)>0$ quantifies the information acquired by the demon before $t=t_0$, and $-\Delta I(t)$ the amount of information consumed in the transformation. Comparing with Eq.~\eqref{entprodj} in the case $j=S$, $j'=D$, one can see that the left-hand side can now take negative values provided $I(t)<I_0$ (some information is consumed). In other words, information acquired about the system, as quantified by the quantum mutual information, can be seen as a thermodynamic resource that the demon can consume to obtain effects akin to a work expenditure. In the most striking case, the demon provides no work nor heat, and still increases the system's free energy, or decreases its entropy \cite{Sanchez19}.}\\

\section{Physical Model}

\red{From now on, we consider the case of a bi-partite system $j,j'=\text{Q},\text{C}$, where Q stands for a qubit (e.g. a two-level atom) and C a harmonic oscillator (hereafter dubbed the cavity), coupled via the Jaynes-Cummings interaction}
\bb
 \hat V_\text{QC} =  i\hbar g ( a \sigma_+ - a^\dagger \sigma_-) ,
\ee
where $a$ denotes the bosonic annihilation operator of the cavity, and $\sigma_-=\ket{g}\bra{e}=\sigma_+^\dagger$ the lowering operator of the qubit. We focus on resonant conditions, where the qubit and cavity Hamiltonians are  $\hat H_\text{Q} = \hbar\omega_0 \sigma_+\sigma_-$ and $\hat H_\text{C}=\hbar\omega_0 a^\dagger a$, respectively. We assume that the cavity is initially prepared in a coherently-displaced thermal state
\bb
 \hat\rho_\text{C}(0) = D(\alpha_0)\hat w_{\beta_C(0)} D^\dagger(\alpha_0) ,
\ee
where $\hat w_{\beta_C(0)} = \frac{e^{-\beta_C(0)\hbar\omega_0 a^\dagger a}}{Z_\text{C}(0)}$ is a thermal equilibrium state at inverse temperature $\beta_C(0)$, with $Z_\text{C}(0) = \text{Tr}\{e^{-\beta_C(0)
\hbar\omega_0 a^\dagger a}\}$ and
\bb
 \bar n = (e^{\beta_C(0)\hbar\omega_0}-1)^{-1}
\ee
\red{is the initial number of thermal excitations in the oscillator}.
$D(\alpha_0) = e^{\alpha_0a^\dagger-\alpha_0^*a}$ denotes the phase-space displacement operator with amplitude
\bb
\alpha_0 = \sqrt{n_0}e^{i\phi_0}.
\ee
We are interested in the limit $n_0\gg \bar n,1$, in which the cavity is expected to induce a classical drive for the qubit. 

\red{It is useful for the following to emphasize that the model verifies $[\hat H_\text{C}+\hat H_\text{Q}, \hat V_\text{QC}]=0$, such that the sum of the qubit and cavity internal energies $E_C(t)+E_Q(t)$ is a constant of motion, and that over any time interval
\bb 
\Delta E_\text{C} = - \Delta E_\text{Q}.
\ee}

\begin{figure}
    \centering
    \includegraphics[width=\linewidth]{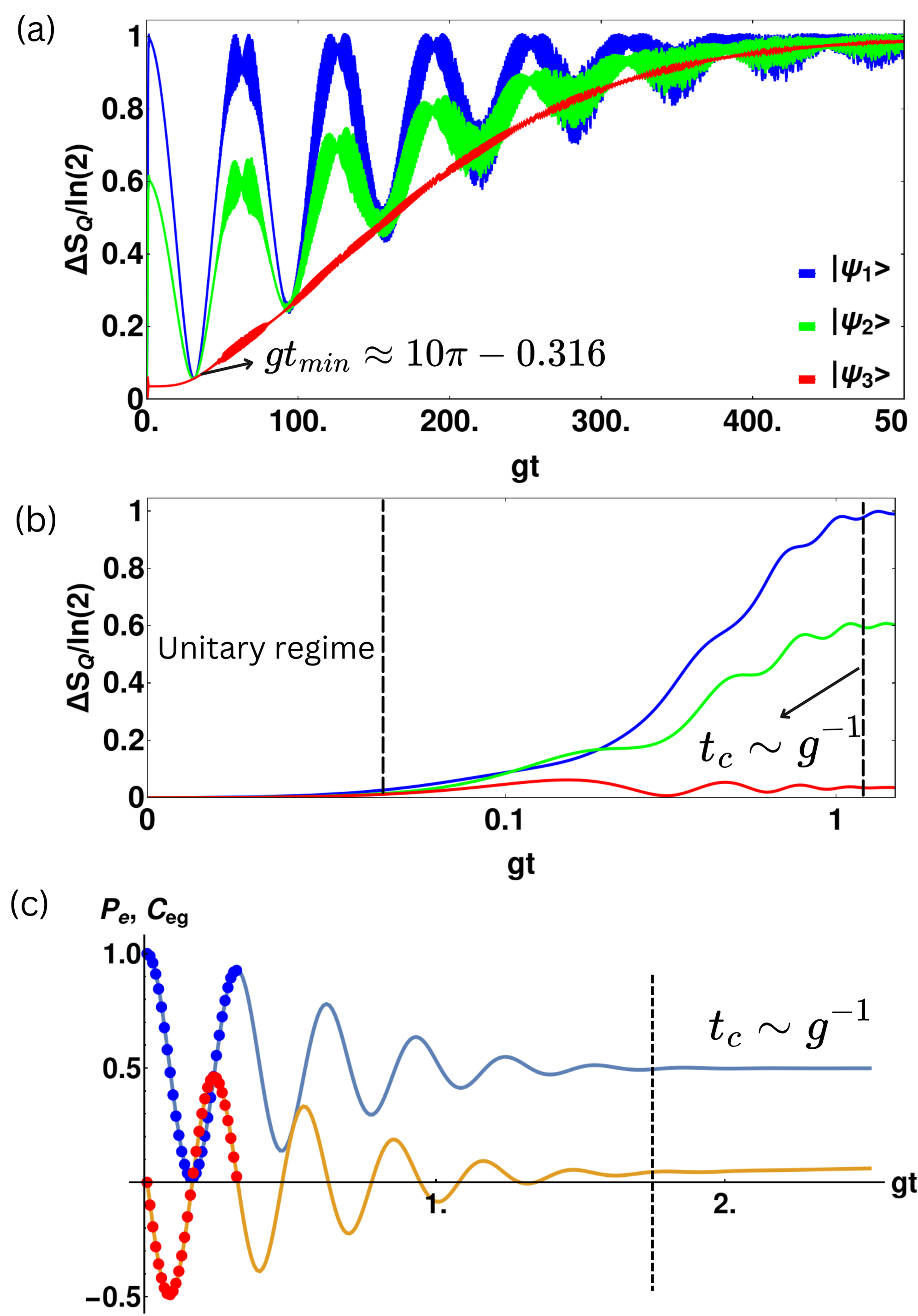}
    \caption{(a) Entropy variation $\Delta S_\text{Q}$ of the qubit over a long time scale for different initial qubit states $\ket{\psi_1} = \ket{e}\bra{e}$, $\ket{\psi_2} =\frac{1}{\sqrt{2}} (\ket{e} + e^{i\frac{\pi}{4}} \ket{g})$, and $\ket{\psi_3} = \ket{+_y}= \frac{1}{\sqrt{2}} (\ket{e} + i \ket{g})$. (b) zoom in the early times of (a), in semilog scale. 
    (c) Excited state population $P_e=\bra{e}\hat\rho_\text{Q}(t)\ket{e}$ (in blue) and coherence $C_{eg}=\bra{e}\hat\rho_\text{Q}(t)\ket{g}$ (in orange and red) of the qubit when it is initialized in $\hat{\rho}_\text{Q}(0) = \ket{e}\bra{e}$. The dots correspond to expansions up to second order in $1/n_0$ and the solid lines are obtained from the numerical evaluation of the exact expressions, truncating the cavity Hilbert space to $N_\text{ph} = 500$. Parameters: $n_0=100$, $\bar n=1$.}
    \label{f:3regimes} 
\end{figure}

\red{Applying the framework introduced in Secs.~\ref{s:Autothermo}-\ref{s:MaxwellDemons} to this model, we find that the dynamics of the autonomous machine features three successive temporal regimes, in which, respectively, the cavity behaves as a quasi-ideal work source (unitary regime), then acquires information about the system (measurement regime), and finally performs autonomous feedback on the qubit. Those three regimes can be identified from the time-evolution of the qubit's von Neumann entropy as shown in Fig. \ref{f:3regimes}(a) and (b). We analyze them in the following sections, and show the cavity behaves as an autonomous Maxwell demon.}\\

\section{Unitary regime}
\begin{figure}
        \centering
\includegraphics[width=\linewidth]{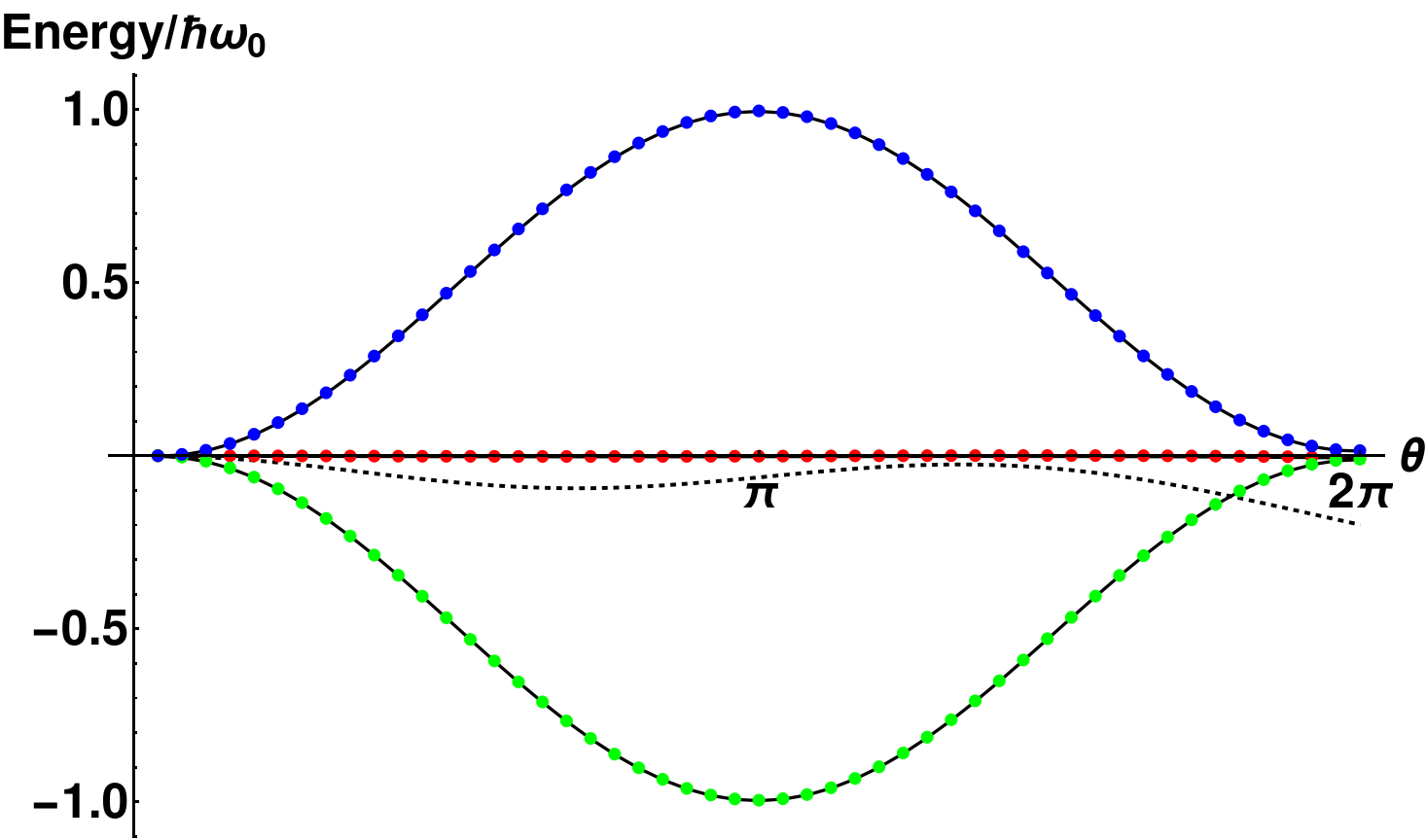}
\caption{
Heat $Q_\text{C}$ (red), work $W_\text{C}$ (green) and internal energy variation of the cavity $\Delta E_\text{C}$ (blue) as a function of $\theta=2gt\sqrt{n_0}$, for $\hat{\rho}_\text{Q}(0) = \ket{e}\bra{e}$, $n_0=500$ and $\bar n=1$. Dots are obtained from the second order expansion in $1/n_0$, and exact expressions with truncated cavity Hilbert space are displayed in solid black (not distinguishable). For smaller values of $n_0$, the heat provided by the cavity is non-negligible, as shown by the dashed black line for $n_0=10$.
}
        \label{f:UnitaryRegime}
    \end{figure}

For times $t \ll  g^{-1}$, the cavity and the qubit density operators remain approximately factorized, such that the cavity behaves as a quasi-ideal source of work inducing a classical drive on the qubit. 
As a consequence, the qubit follows a unitary evolution ruled by the effective Hamiltonian
\begin{equation}
    \begin{aligned}
       \hat H^{\text{eff}}_\text{Q}(t) &= \text{Tr}_\text{C}\{\hat V_\text{QC}\hat{\rho}_\text{C}(t)\}\\
      &=i\hbar g (\moy{a(t)}\sigma_+-\langle{a^\dagger(t)}\rangle\sigma_-).
    \end{aligned}
\end{equation}
The latter induces Rabi oscillations with an approximately constant frequency $\Omega \simeq g\vert \! \moy{a(0)} \! \vert=g\sqrt{n_0}$ (see Fig.~\ref{f:3regimes}(c) and Appendix~\ref{s:worksource}).  To identify thermodynamic evidence of the work source behavior of the cavity, and deviations from it, we plot in Fig.~\ref{f:UnitaryRegime} the work $W_\text{C}$ and the heat $Q_\text{C}$ provided by the cavity to the qubit, together with the variation of the cavity's internal energy $\Delta E_\text{C}$. We see that for large enough values of $n_0$, the cavity mostly provides work and almost no entropy is produced during a single Rabi oscillation. We formalize this result by expanding as a power series of $1/n_0$ the exact cavity and qubit state in the interaction picture
\bb
\hat\rho(t) = \hat U(t)(\hat\rho_\text{Q}(0)\otimes\hat\rho_\text{C}(0))\hat U^\dagger(t) ,
\ee
where the analytical expression of $\hat{U}(t)$ is provided in Appendix~\ref{s:Model}. For $\hat\rho_\text{Q}(0)=\ket{e}\bra{e}$, the qubit population $P_e(t) = \bra{e}\hat\rho_\text{Q}(t)\ket{e}$ and coherence $C_{eg}(t) = \bra{e}\hat\rho_\text{Q}(t)\ket{g}$ read:
\bb
P_e(t) &=& \cos^2{\frac{\theta}{2}} + \frac{ \delta P_{e}(\theta,\bar{n})}{n_0} +  {\cal O}\!\left(\!\frac{1}{n_0^2}\!\right)\nonumber\\
C_{eg}(t)&=&-e^{i\phi_0} \left[ \frac{1}{2} \sin{\theta} + \frac{\delta C_{eg}(\theta,\bar{n})}{n_0} \right]  + {\cal O}\!\left(\!\frac{1}{n_0^2}\!\right)\!,\quad\;\;
\ee
where $\theta = 2\Omega t = 2gt \sqrt{n_0}$ corresponds to the rotation angle in the Bloch sphere along an axis of the equator with azimuthal angle $\phi_0$. The first order corrections to an ideal unitary dynamics are given by
\begin{equation}
    \begin{aligned}
       \delta P_e\!\left(\theta, \Bar{n}\right) \!&=&\!\!\!\!\!\! - \frac{\theta}{16} \left( \left(1+2\Bar{n}\right) \theta \cos{\theta} + \left(3+2\Bar{n} \right)\sin{\theta} \right)\\
\delta C_{eg}\!\left(\theta,\Bar{n}\right) \!&=&\!\!\! \!\!\! \frac{-2\theta\!\!+\!\! \left(3\!+\! 2\bar{n}\right)\!\theta\! \cos\!\theta\!-\!\left(1\!+\!2\bar{n}\right)\!\!\left(\!1\!+\! \theta^2\right)\!\sin\!\theta}{ 16}.
    \end{aligned}
\end{equation}\\
These first-order truncations show good agreement with exact numerical results over time-scales much shorter than $g^{-1}$, as shown in Fig.~\ref{f:3regimes}(c). In addition, our expansion shows that the amount of exchanged heat
\begin{equation}
    \begin{aligned}
       Q_\text{C}\!\left(t\right)\!=\! - \frac{\hbar \omega_0}{16 n_0}&\!\left(\theta^2\!+\! \sin^2{\theta}\!+\!2 \left(1+2\bar{n}\right)\!\theta \sin\!{\theta}  \right)\\
       \!&+\! {\cal O}\left(\frac{1}{n_0^2}\right)
    \end{aligned}
\end{equation}
scales as $1/n_0$ (see Fig.~\ref{f:UnitaryRegime}), confirming that the cavity tends to an ideal work source behavior in the limit $n_0\to\infty$. Similarly, entropy production associated with the qubit's transformation is found to \red{vanish} for large $n_0$ (see Appendix \ref{s:worksource}).


\section{Measurement-induced decoherence}
It is known from previous studies \cite{meystre1973destruction,narozhny1981coherence, GeaBanacloche90}, as well as measurements on Rydberg atoms in cavities \cite{rempe1987observation, Auffeves03}, that the qubit decoheres, and Rabi oscillations collapse, over a time scale $t_c \sim g^{-1}$ (see Fig.~\ref{f:3regimes}(c)). This decoherence is due to the evolution of the field towards two orthogonal states, with macroscopically different values of the complex field amplitudes $\moy{a(t)}$ conditioned on the state of the qubit. This evolution can be interpreted as \red{an autonomous ``measurement''} of the qubit by the cavity~\footnote{A more correct terminology here would be a ``pre-measurement". The measurement process would be complete, and the result would truly become an objective fact, only if it were copied multiple times into the state of a large number of different environmental degrees of freedom  \cite{Zurek09}. If the cavity and qubit continue to evolve in isolation, revivals of the Rabi oscillations would eventually be observed \cite{GeaBanacloche90}. However, the measurement analogy remains fully consistent on the time interval considered in this article.} \cite{haroche1998entanglement,GeaBanacloche91}. 

The measurement basis is set by the eigenbasis of $\hat H_\text{Q}^\text{eff}(0)$, and therefore by the initial phase of the field. Explicitly, it reads $\{\cos(\frac{\phi_0}{2})\ket{+_y} -i \sin(\frac{\phi_0}{2})\ket{-_y},-i \sin(\frac{\phi_0}{2})\ket{+_y}+\cos(\frac{\phi_0}{2})\ket{-_y}\}$, where $\ket{\pm_y}$ are the eigenstates of $\sigma_y = i(\sigma_--\sigma_+)$. It therefore corresponds to a measurement along the rotation axis of the unitary regime, which lies in the equator of the Bloch sphere. 
In the case $\phi_0=0$, which we consider hereafter, the qubit is measured along the $y$-axis. The total state verifies for $t\lesssim t_c= {\cal O}(g^{-1})$ \red{(see Appendix \ref{s:Meas}):}

\bb\label{eq:rhoCnumu}
\hat\rho(t) =\!\!\! \sum_{\nu,\mu=\pm}\rho_{\nu\mu}  \ket{\nu_y}\bra{\mu_y} \otimes \hat{\rho}_\text{C}^{\nu\mu}(t) + {\cal O}\!\left(\frac{1}{n_0}\right)\!,\;\;\;
\ee
where $\rho_{\nu\mu} =  \bra{\nu_y}\rho_\text{Q}(0)\ket{\mu_y}$ are the coefficients of the qubit density operator in the measurement basis and
\bb
\hat{\rho}_\text{C}^{\nu\mu}(t) \!\!\!&=&\!\!\! e^{\nu i \chi \delta_{\nu\mu}} D(\sqrt{n_0} e^{\nu \frac{igt}{2\sqrt{n_0}}}) \hat{w}_{\beta_C(0)} D^\dagger(\sqrt{n_0} e^{\mu \frac{igt}{2\sqrt{n_0}}}),\;\;\;\;
\ee
are conditional operators in the cavity space with $\chi = gt(\sqrt{n_0} + 1/\sqrt{n_0})$ and $\delta_{\nu\mu}$ the Kronecker delta.  The diagonal terms $\hat\rho_\text{C}^{\nu\nu}$ are conditional cavity states associated with measurement results $\nu = \pm$, and are characterized by field amplitudes
\bb
\alpha_\nu(t) = \sqrt{n_0} e^{\nu \frac{igt}{2\sqrt{n_0}}}.\label{eq:alphanu}
\ee
For $t\in[0,t_c]$, they fulfill
\bb
\alpha_\nu(t)= \sqrt{n_0} +i\nu\frac{gt}{2}+{\cal O}\left(1/\sqrt{n_0}\right),
\ee
and therefore differ by the sign of their purely imaginary component. For $t\sim g^{-1}$, those two field amplitude become fully distinguishable (correspond to orthogonal cavity states). The trace of the cross terms $\text{Tr}\{\hat\rho_\text{C}^{+-}(t_c)\}$ vanishes for $n_0\gg \bar{n},1$, which implies maximal qubit-cavity correlations, and hence a complete measurement \cite{WisemanBook}. In particular, the reduced qubit state then verifies
\bb
\hat\rho_\text{Q}(t_c) = \rho_{++}\ket{+_y}\bra{+_y} + \rho_{--}\ket{-_y}\bra{-_y}+{\cal O}(\frac{1}{n_0}) ,
\ee
achieving complete dephasing of the Rabi oscillations in the classical limit. \\

\section{Autonomous feedback and qubit purification}
\begin{figure}
    \centering
    \includegraphics[width=0.95\linewidth]{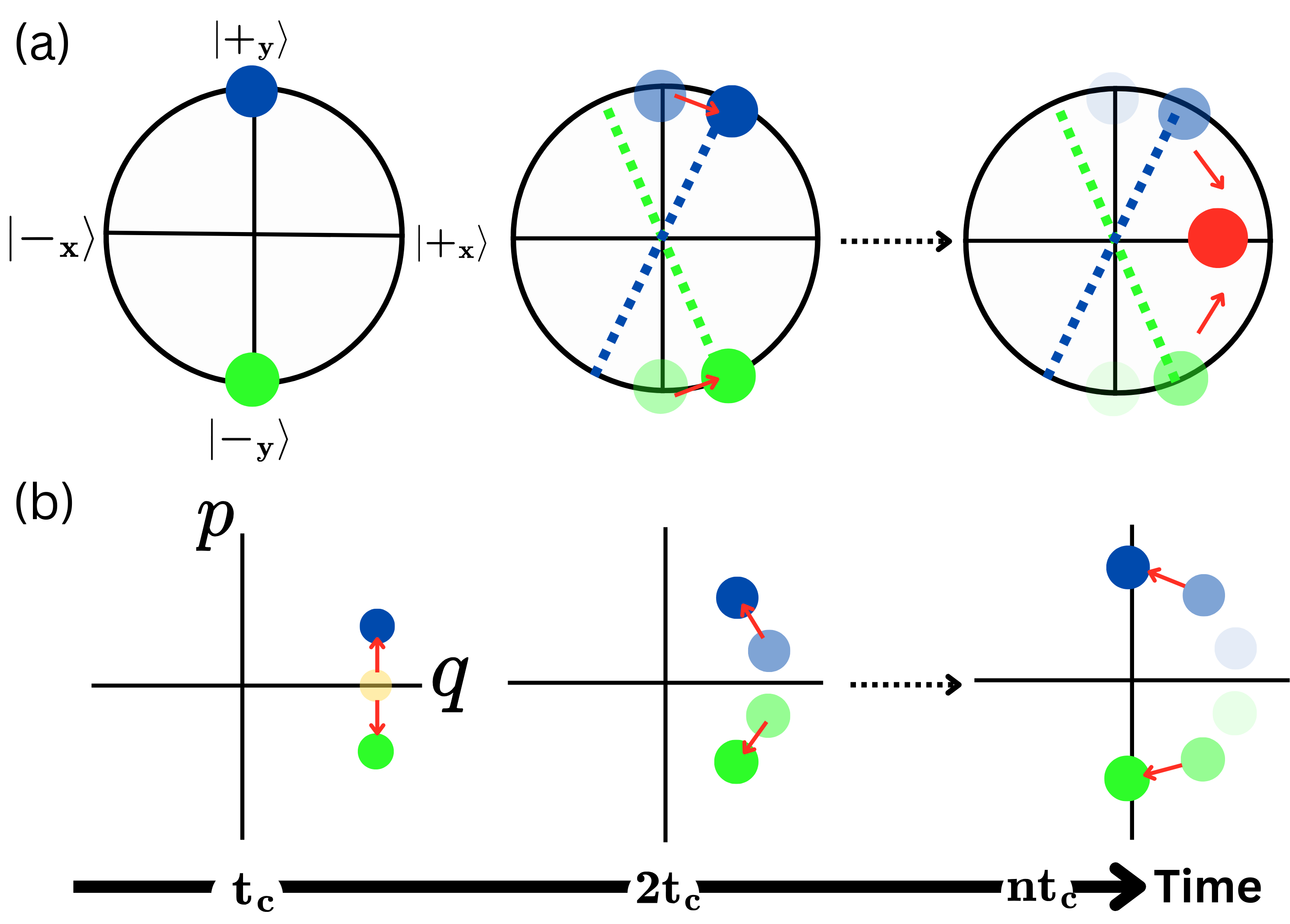}
    \caption{Autonomous feedback mechanism. (a) Trajectory of the qubit  on the equator of the Bloch sphere starting from states $\ket{+_y}$ (blue) and $\ket{-_y}$ (green) and (b) of the average field amplitude. The red circle locates the qubit state reached at $t_\text{min}$. The yellow circle locates the initial cavity displacement, for $\phi_0=0$.}
    \label{f:feedback}
\end{figure}

Remarkably, the behavior of the machine presents a third regime on an even longer timescale $t \sim \sqrt{n_0} g^{-1}$, in which the cavity induces an effective drive on the qubit depending on the result of the measurement in the $\{\ket{\pm_y}\}$ basis. This evolution therefore constitutes an example of autonomous feedback. This feedback is able to purify any initial qubit state when $n_0$ is sufficiently large, as shown for zero initial cavity temperature in Ref.~\cite{GeaBanacloche91}\red{, and here extended to an initially displaced thermal state}. We first note that at $t\sim t_c$, the cavity field $\text{Tr}\{a\hat\rho_\text{C}^{\nu\nu}\} =\alpha_\nu(t)$ depends on the measurement result $\nu$. It is instructive to analyze the joint qubit-cavity dynamics along the two independent branches associated with the qubit found in $\ket{\pm_y}$. As, \emph{inside each branch}, the qubit-cavity state is factorized up to ${\cal O}(1/n_0)$ corrections, one can again consider the cavity as an ideal work source, inducing a classical drive on the qubit of the form (see Appendix~\ref{s:Feedback})
\bb
\hat H^{\text{eff}(\nu)}_\text{Q}(t) = i\hbar g(\alpha_\nu(t)\sigma_+-\alpha_\nu^*(t)\sigma_-) ,
\ee
where, unlike early times ($t < t_c$), the effective driving Hamiltonian depends on the measurement result $\nu$. At $t_c$, the two Hamiltonians almost coincide with $\hat H_\text{Q}^\text{eff}(0)$, and the qubit measured state is aligned with the field, whatever $\nu$. Then, two different field axes emerge. From Eq.~\eqref{eq:alphanu}, we see that they both lie in the equator and rotate in opposite directions with angular velocities $\pm\frac{g}{2\sqrt{n_0}}$ much smaller, by a factor of $1/n_0$, than the energy splitting of $\hat H^{\text{eff}(\nu)}_\text{Q}(t)$, still given by the Rabi frequency $\Omega$. As a consequence, the qubit follows adiabatically the rotation axis inside each measurement branch (see Fig.~\ref{f:feedback}). The two field axes meet again at $t=t_\text{min} = \sqrt{n_0}\pi g^{-1}$ where they overlap with the $x$-axis of the Bloch sphere, driving in both branches the qubit to state $\ket{+_x}$. As a consequence of this autonomous feedback, an arbitrary initial qubit state $\hat\rho_\text{Q}(0)$ is deterministically purified and brought to $\ket{+_x}$ 
(or generally for a phase $\phi_0$ to state $(1/\sqrt 2)\left(\ket{g}+e^{i\phi_0}\ket{e}\right)$) at time $t_\text{min}$. 

It is worth noting that this ideal purification occurs in the limit $n_0\to \infty$, such that $t_\text{min}$ also diverges. Nevertheless, our expansion at first order in ${\cal O}(1/n_0)$ allows us to identify corrections for finite $n_0$. Mainly, the corrections are due to correlations between the cavity and the qubit inside each branch, which are responsible for a purity loss of the qubit state. As a result, the qubit achieves a non-zero minimum entropy (scaling as ${\cal O}(1/n_0)$), at a time $t_\text{min}$ which slightly differs from $\sqrt{n_0}\pi g^{-1}$ (See Fig.~\ref{f:3regimes}(a)). In summary, there is a trade-off between the purity of the attained qubit state and the time needed to purify the qubit. 

As seen in Fig.~\ref{f:3regimes}(a), \red{after $t_\text{min}$}, the entropy of the qubit goes through a series of minima with a decreasing purity at times 
\bb
t_k\simeq (2k-1)\sqrt{n_0}\pi,\quad(t_1\equiv t_\text{min}),
\ee
\red{separated by maxima of increasing entropy. This behavior can be attributed to series of measurement and feedback sequences over the time intervals $[t_{k-1},t_k]$, with degraded performances due to the accumulation of qubit-cavity correlations during the intra-branch dynamics, at finite $n_0$. The entropy maxima are reached after each measurement step. When the qubit starts already in one of the two pointer states $\ket{\pm_y}$ (red curve), the measurement has no effect and the cavity-qubit correlations associated with the measurement do not build up, hence the absence of maxima. In that case, the qubit entropy increase is only due to intra-branch correlations}.  After a few oscillations, as intra-branch correlations keep growing, the qubit entropy saturates at $\log 2$, the maximum physical value for a qubit.

\section{Thermodynamic signature of an autonomous Maxwell demon}
\begin{figure}[h]
        \centering
        \includegraphics[width=\linewidth]{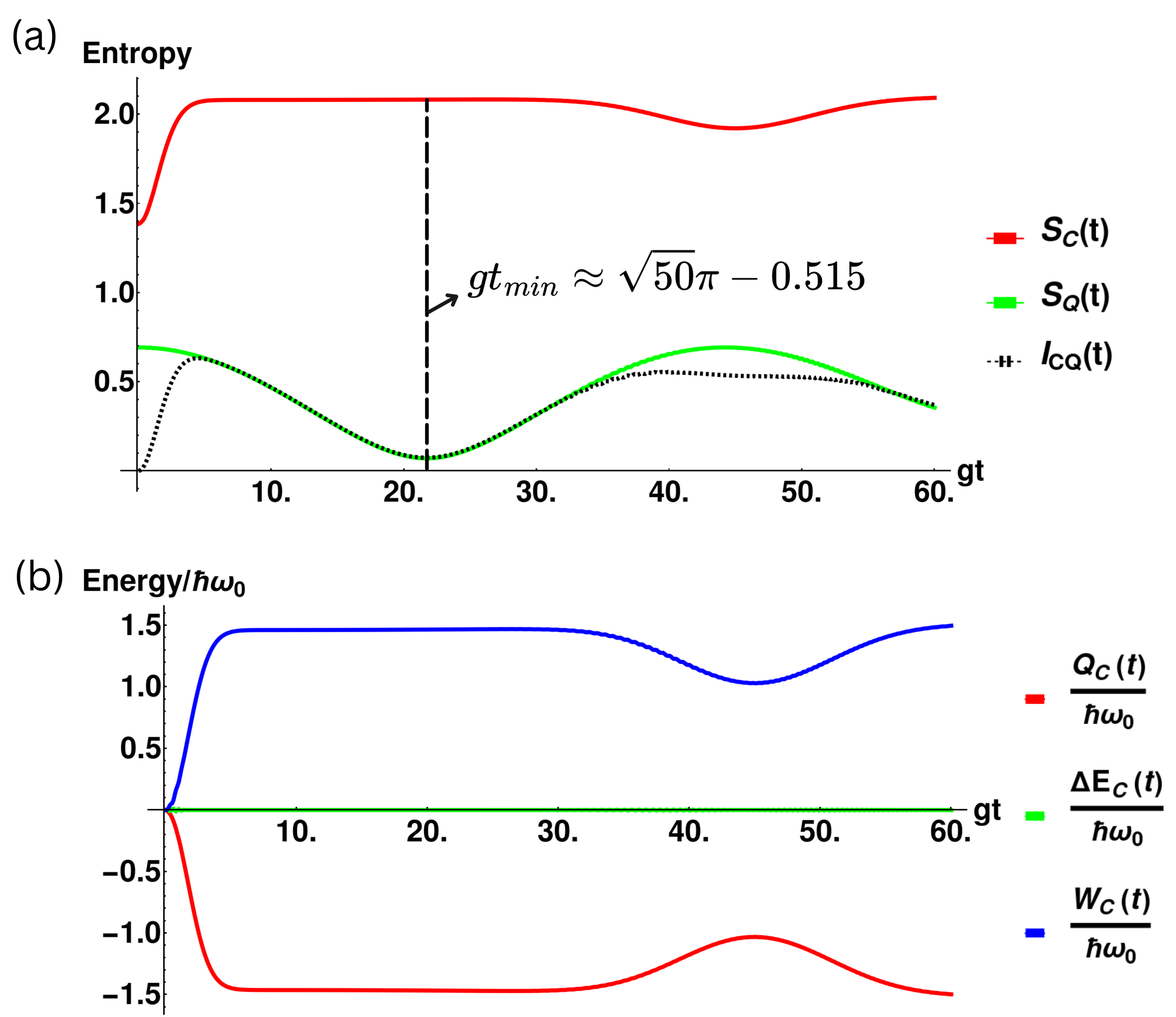}
        \caption{(a) von Neumann entropy of the cavity $S_\text{C}$ and of the qubit $S_\text{Q}$, and mutual information between the qubit and cavity $I_\text{QC}$. (b) Cavity heat $Q_\text{C}$, work $W_\text{C}(t)$ and internal energy variation $\Delta E_\text{C}(t)$ \red{associated to the time interval $[0,t]$}. For both (a) and (b), $n_0=50$, $\bar n=1$. Plotted quantities are computed from exact calculation with truncated cavity Hilbert space.  }
        \label{f:demon}
    \end{figure}

We now use our thermodynamic framework to analyze the purification process, and show that \red{the cavity behaves as an autonomous Maxwell demon. We assume in this last section that the qubit is initially prepared in the maximally mixed state $\rho_\text{Q}(0)= \idop/2$ (center of the Bloch sphere)}. Fig.~\ref{f:demon} shows the evolution of the qubit and cavity entropies, together with their quantum mutual information
\bb
I_\text{QC}(t) = S_\text{C}(t) + S_\text{Q}(t) - S_\text{QC}(t).
\ee 
We also plotted the work $W_\text{C}(t)$ and heat $Q_\text{C}(t)$ provided by the cavity, together with its internal energy variation $\Delta E_\text{C}$. \red{By comparing the initial and final states of the qubit (respectively, $\rho_\text{Q}(0)$ and $\ket{+_x}\bra{+_x}$ in the large $n_0$ limit), we first note that the overall effect on the qubit is a net decrease of its entropy from $\log 2$ to 0, without changing its energy. 

As it can be seen from the overall evolution of the cavity, this entropy decrease is made possible by an expense of work performed by the cavity on the qubit, as in a 
Landauer reset protocol \cite{Parrondo15}. 
One can use Eq.~\eqref{entprodj} in the case $j=Q$ and $j'=C$, and for the time interval $[0,t_\text{min}]$ together with the fact that $\Delta E_C=-\Delta E_Q=0$ (in the large $n_0$ limit), to show that $-\beta_C(0) Q_C(t_\text{min})= \beta_C(0) W_C(t_\text{min})\geq - \Delta S_Q$. The Landauer bound $W_C(t_\text{min}) \geq  \beta_C^{-1}(0)\log 2$ is therefore retrieved when $n_0\gg 1$ such that the purification is complete $\Delta S_Q\simeq-\log 2$. We emphasize however that, here, the reset protocol is not occurring via a deterministic driving in presence of a heat source, as typically considered, but rather via the autonomous feedback mechanism induced by the cavity}.

\red{To highlight the signatures of a Maxwell demon, we analyze separately the two time intervals $[0,t_c]$ (information acquisition) and $[t_c,t_\text{min}]$ (purification). We build on the dynamical analysis above to interpret the evolution of the thermodynamic quantities along those two steps. We refer to Fig.~\ref{f:demon} where are plotted the numerical evaluations of exact expressions for the relevant information theoretic and thermodynamic quantities.} 
\red{During time interval $[0,t_c]$ ($gt \in [0,1]$), the qubit-cavity quantum mutual information grows from $0$ (as expected for an initial factorized state) to almost its maximal value $I(t_c)\simeq \log 2$ reached around $t_c$ (see Fig.~\ref{f:demon}a and Appendix~\ref{s:ThermoDemon} for an analytical derivation in the large $n_0$ limit). This step corresponds to the transfer of information about the qubit into the demon's memory. The reduced entropy of the cavity also increases by the same amount. In the considered case of a qubit starting in the center of the Bloch sphere $\rho_Q(0)=\idop/2$, no Rabi oscillation occur, and the qubit remains in state $\rho_Q(0)=\idop/2$. As a consequence, the qubit's energy, and therefore the cavity's energy as well, are constant. The time interval $[0,t_c]$ is then associated to a work-to-heat transfer inside the cavity, as seen from Fig.~\ref{f:demon}b. }

\red{During time interval $[t_c, t_\text{min}]$ ($gt \in [1,gt_\text{min}]$), the qubit is purified $S_Q(t_\text{min}) \simeq 0$ (in the large $n_0$ limit), while its energy is conserved (its state stays within the equatorial plane of the Bloch sphere on the whole interval, as discussed in Section \ref{s:Feedback}). Along this time interval as well, the cavity energy is therefore conserved. As the cavity entropy is also constant (see Appendix \ref{s:ThermoDemon}), the cavity provides no work nor heat to the qubit in the time interval $[t_c, t_\text{min}]$. This leads to $\Delta S_Q -\beta_C(t_c)Q_C = \Delta S_Q < 0$, which constitutes an apparent violation of Clausius inequality, i.e. the signature of the Maxwell demon. Consistency with the second law is restored by taking into account the resource being consumed, as done in Eq.~\eqref{entproddemon}, that is, the quantum mutual information between the cavity and the qubit which simultaneously decreases from $\log 2$ to 0. Those results are analytically confirmed by our perturbative calculation in the large $n_0$ limit (see Appendix \ref{s:ThermoDemon}). We note that the Gaussian approximation used in the unitary regime does not apply, even at dominant order, as the reduced cavity state is a mixture of two orthogonal distributions in phase space, which is highly non-Gaussian. An interesting outcome of the analytical calculation is to show that the inequality \eqref{entproddemon} is saturated in our model in the large $n_0$ limit (see Appendix \ref{s:ThermoDemon}), meaning a reversible conversion of mutual information into entropy decrease. }

\red{As it can be seen in Fig.~\ref{f:demon}, when $n_0$ is finite, the qubit entropy still decreases, albeit to a finite value, and the cavity still exchanges no work nor heat during the interval $[t_c, t_\text{min}]$. The entropy decrease of the qubit remains entirely fueled by the consumption of quantum mutual information. We find that such decrease of the qubit's entropy persists for small values of $n_0$, down to a threshold of approximately $n_0\sim 5$. Below, the qubit entropy starts exhibiting a non-monotonous evolution and the measurement and feedback steps are not clearly marked anymore.}


\red{Thus, we can conclude that a behavior of autonomous Maxwell demon emerges from the Jaynes-Cummings interaction, in the case of an initially coherently-displaced cavity, and is naturally captured by our thermodynamic framework. }

\red{We finish by a few comments to discriminate the mechanism of autonomous Maxwell demon presented here from other spontaneous purification processes. Purity minima are expected to occur in any quantum system initialized in a pure state and coupled to another finite-size system. Indeed, Poincaré recurrences \cite{Bocchieri57,Schulman78} ensure that the total system will repeatedly reach states arbitrarily close to its initial state. In contrast, here, the initial state of the qubit is \emph{not restored} when its purity reaches its minima. Instead, the pure state it reaches is entirely determined by the cavity's phase, and is independent of the qubit's initial state. Consequently, the qubit is purified even if it was not initially in a pure state. Besides, an entropy decrease could also be induced in the qubit by simply coupling it to another system of low entropy, e.g. another qubit initially in a pure state. In contrast here, the cavity can have initially an entropy comparable to, or even higher than that of the qubit, while still enabling perfect purification, provided its coherent amplitude is large enough. This is possible owing to the infinite Hilbert space of the cavity, able to perfectly store the result of the measurement, and conserve it while performing a unitary feedback on the qubit. Another pre-requisite is the non-thermal nature of the initial cavity state, which makes it suitable to provide work \cite{Elouard23}. Finally, notice that a large Hilbert space is also required for the ideal work source behavior observed at short times: more precisely, the work source Hilbert space needs to support quantum states of average energy differing by the qubit transition energy $\hbar\omega_0$, while also having an overlap arbitrarily close to $1$. This combination of properties, specific to infinite-dimensional Hilbert spaces, is required to generate coherent, purity preserving, Rabi oscillations of the qubit, despite the dynamical constraint of conserving the cavity and qubit total energy \cite{Stevens22,Rogers22}.}

\section{Conclusions}
We have analyzed the dynamics of a qubit coupled to a cavity prepared in a displaced coherent state of very large displacement amplitude, a state commonly considered as guaranteeing a classical drive behavior of the cavity. Owing to our framework for the thermodynamics of autonomous machines, we have shown that at short enough times, the cavity indeed behaves as an ideal work source inducing a time-dependent drive on the qubit. However, over a time-scale set by the qubit-cavity interaction strength, the cavity performs a measurement of the qubit state in a basis set by the initial cavity displacement phase, responsible for the collapse of the Rabi oscillations. We have shown that this measurement is followed by an autonomous feedback, a state-dependent drive able to purify the qubit. From the analysis of thermodynamic quantities, we relate this purification to the cavity behaving as an autonomous Maxwell demon, spending work to build mutual information with the qubit, and then consuming this mutual information \red{to} decrease the qubit's entropy. %
Those results demonstrate the potential of consistent thermodynamic frameworks \red{suitable for autonomous systems}, like the one of Ref.~\cite{Elouard23}, to shed new light on complex quantum dynamics. 
The parameters needed to implement our scheme are within state-of-the-art, e.g. in superconducting qubit-cavity systems, where strong couplings with respect to damping rates can be achieved.
Natural extensions of this work include optimizing the initial cavity state to improve the purification process, e.g. by attaining purer states at a fixed feedback time, and testing the robustness of the purified qubit state to cavity preparation and residual qubit-cavity detunings. 

\section*{Acknowledgements} 
C.E. acknowledges funding from French National Research Agency (ANR) under project ANR-22-CPJ1-0029-01. We thank Alexia Auffèves for fruitful discussions.


\bibliographystyle{unsrt}

\onecolumngrid
\appendix

\section{Physical Model}\label{s:Model}
We consider a qubit $\left(Q\right)$ interacting with a harmonic oscillator ($C$) through a resonant  Jaynes-Cummings coupling. The total Hamiltonian is given by 
\begin{equation}
    \begin{aligned}
       \hat{H} &= \hat{H}_\text{C} + \hat{H}_\text{Q} + \hat{V}_\text{QC} 
               = \hbar \omega_0 a^\dagger a  + \hbar \omega_0 \sigma_+ \sigma_- + i\hbar g \left(a \sigma_+ - a^\dagger \sigma_-\right) 
               \label{eq:Model}
    \end{aligned}
\end{equation}
with $\omega_0$ the common transition frequency  and  $g$ the coupling constant between the qubit and the cavity. The operators $a$ and $a^\dagger$ are the usual bosonic annihilation and creation operators and $\sigma_+ = \ket{e}\bra{g} = \sigma_-^\dagger$ with $\ket{e}$ and $\ket{g}$ the excited and ground states of the qubit, respectively.

Due to the resonance condition, the evolution of the total system in the interaction picture is governed by Hamiltonian $\hat H_I(t) = \hat{V}_\text{QC}$. We deduce the interaction picture evolution operator $\hat{U}(t) = e^{-\frac{it}{\hbar} \hat{V}_\text{QC} }$, which can expanded as \cite{GeaBanacloche90}:

\begin{equation}
    \hat{U}\left(t\right) = \cos\left(g t \sqrt{a^{\dagger}a + 1}\right) \ket{e}\bra{e} + \cos\left(g t \sqrt{a^{\dagger}a}\right) \ket{g}\bra{g}  + \frac{\sin\left({g t \sqrt{a^{\dagger}a + 1}}\right)}{\sqrt{a^{\dagger}a + 1}} a \sigma_{+} - \frac{\sin\left({g t \sqrt{a^{\dagger}a }}\right)}{\sqrt{a^{\dagger}a }} a^{\dagger} \sigma_{-} \label{eq:genunitary}
\end{equation}

\section{Quasi-ideal work source behavior of the cavity}
\label{s:worksource}
To verify the quasi-ideal work source behavior of the cavity, we consider the following initial factorized state for the qubit and cavity
\begin{equation}
    \begin{aligned}
       \hat{\rho}\left(0\right) = \ket{e}\bra{e} \otimes D\left(\alpha_0\right) \hat{w}_{\beta_C\left(0\right)} D^\dagger\left(\alpha_0\right), \hspace{5pt} \alpha_0 = \sqrt{n_0}e^{i \phi_0} \label{eq:initstate} \; . 
    \end{aligned}
\end{equation}
Under the unitary evolution the state $ \hat{\rho}(t) = 
\hat{U}(t) \hat{\rho}\left(0\right) \hat{U}^\dagger(t)$ takes the form
\begin{equation}
    \begin{aligned}
        \hat{\rho}\left(t\right)
        &= \ket{e}\bra{e} \otimes \cos\left(g t \sqrt{a^{\dagger}a +1}\right) D\left(\alpha_0\right) \hat{w}_{\beta_C\left(0\right)} D^\dagger\left(\alpha_0\right) \cos\left(g t \sqrt{a^{\dagger}a + 1}\right) \\ -
    & \ket{e}\bra{g} \otimes \cos\left(g t \sqrt{a^{\dagger}a + 1}\right) D\left(\alpha_0\right) \hat{w}_{\beta_C\left(0\right)} D^\dagger\left(\alpha_0\right)  a \frac{\sin\left({g t \sqrt{a^{\dagger}a }}\right)}{\sqrt{a^{\dagger}a }} \\ 
    &- \ket{g}\bra{e} \otimes \frac{\sin\left({g t \sqrt{a^{\dagger}a }}\right)}{\sqrt{a^{\dagger}a }} a^{\dagger} D\left(\alpha_0\right) \hat{w}_{\beta_C\left(0\right)} D^\dagger\left(\alpha_0\right) \cos\left(g t \sqrt{a^{\dagger}a + 1}\right)  \\ 
    &+ \ket{g}\bra{g} \otimes  \frac{\sin\left({g t \sqrt{a^{\dagger}a }}\right)}{\sqrt{a^{\dagger}a }} a^{\dagger} D\left(\alpha_0\right) \hat{w}_{\beta_C\left(0\right)} D^\dagger\left(\alpha_0\right) a \frac{\sin\left({g t \sqrt{a^{\dagger}a }}\right)}{\sqrt{a^{\dagger}a }} \; . \label{eq:eeevolution}
    \end{aligned}
\end{equation}
Tracing over the cavity Hilbert space, the reduced qubit state is given by
\begin{equation}
    \begin{aligned}
\hat{\rho}_\text{Q}\left(t\right) = \begin{bmatrix}
    P_{e}\left(t\right) & C_{eg}\left(t\right) \\
    C_{eg}\left(t\right) & 1-P_{e}\left(t\right)
\end{bmatrix}
\label{eq:rhoQ}
    \end{aligned}
\end{equation}
with population $P_e$ and coherence $C_{eg}$
\begin{equation}
    \begin{aligned}
         P_{e}\left(t\right) &= \sum_n \cos^2\left({g t  \sqrt{n+1}}\right) \bra{n}D\left(\alpha_0\right)\hat{w}_{\beta_C\left(0\right)}D^{\dagger}\left(\alpha_0\right)\ket{n}\\
         C_{eg}\left(t\right) &= -\sum_n \cos\left({g t  \sqrt{n+2}}\right) \sin\left({g t \sqrt{n+1}}\right) \bra{n+1}D\left(\alpha_0\right)\hat{w}_{\beta_C\left(0\right)}D^{\dagger}\left(\alpha_0\right)\ket{n} \; . 
         \label{eq:rhoQPC}
    \end{aligned}
\end{equation}
For $n_0 \gg 1$  the terms $\bra{n}D\left(\alpha_0\right)\hat{w}_{\beta_C\left(0\right)}D^{\dagger}\left(\alpha_0\right)\ket{n}$ and $\bra{n+1}D\left(\alpha\right)\hat{w}_{\beta_C\left(0\right)}D^{\dagger}\left(\alpha\right)\ket{n}$ as a function of $n$ are close to Gaussian distributions peaked around $n_0$ with variance much smaller than the mean. This allows us to expand the trigonometric functions in $P_{e}\left(t\right)$ and $C_{eg}\left(t\right)$ around $n_0$. In the following, we parameterize the population and coherence of qubit in terms of rotation angle $\theta = 2 gt \sqrt{n_0}$ in the Bloch sphere.

For large $n_0$:
\begin{equation}
    \begin{aligned}
       P_e\left(t\right)  &= \sum_n \cos^2\left({\frac{\theta}{2} \sqrt{1 + \frac{n-n_0+1}{n_0}}}\right) \bra{n}D\left(\alpha_0\right)\hat{w}_{\beta_C\left(0\right)}D^{\dagger}\left(\alpha_0\right)\ket{n} \\ 
       &=  \sum_n [ \cos^2{\frac{\theta}{2}} - \frac{\theta \sin{\theta}}{4} \left(\frac{n-n_0+1}{n_0}\right) - \frac{\theta \left(\theta \cos{\theta} - \sin{\theta}\right)}{16} \left(\frac{n-n_0+1}{n_0}\right)^2  + O\left(\frac{1}{n_0^3}\right) ]  \\
       &\times \bra{n}D\left(\alpha_0\right)\hat{w}_{\beta_C\left(0\right)}D^{\dagger}\left(\alpha_0\right)\ket{n}\\
       &=  \cos^2{\frac{\theta}{2}} - \frac{\theta\left( \left(1+2\Bar{n}\right) \theta \cos{\theta} + \left(3+2\Bar{n} \right)\sin{\theta} \right)}{16 n_0} - \frac{\theta \left(1+\Bar{n}\right)\left(1 + 2\Bar{n}\right)\left(\theta \cos{\theta} - \sin{\theta}\right)}{16n_0^2} + O\left(\frac{1}{n_0^3}\right)  \\
       &=  \cos^2{\frac{\theta}{2}} + \frac{\delta P_e\left(\theta, \Bar{n}\right)}{n_0} + O\left(\frac{1}{n_0^2}\right) 
       \label{eq:ExpPe}
    \end{aligned}
\end{equation}
where 
\begin{equation}
    \begin{aligned}
\delta P_e\left(\theta, \Bar{n}\right) = - \frac{\theta}{16} \left( \left(1+2\Bar{n}\right) \theta \cos{\theta} + \left(3+2\Bar{n} \right)\sin{\theta} \right)\; . 
\end{aligned}
\end{equation}
In a similar way
\begin{equation}
    \begin{aligned}
       C_{eg}\left(t\right)  &= -e^{i\phi_0}[ \frac{1}{2} \sin{\theta} + \frac{-2\theta + \left(3 + 2\bar{n}\right) \theta \cos \theta - \left(1 + 2\bar{n}\right)\left(1 + \theta^2\right) \sin \theta}{16 n_0} + \\
 &+ \frac{\left(5+8 \Bar{n}\right)2\theta-\left(11+4 \Bar{n} \left(11+9 \Bar{n}\right)\right) 2\theta \cos{\theta}+\left(12+72 \Bar{n} \left(1+\Bar{n}\right)-\left(9+8 \Bar{n} \left(4+3\Bar{n}\right)\right)
       \theta^2\right) \sin{ \theta}}{64 n_0^2}\\
 &+ O\left(\frac{1}{n_0^3}\right) ]\\
 &= - e^{i\phi_0}[ \frac{1}{2} \sin{\theta} + \frac{\delta C_{eg}\left(\theta,\Bar{n}\right)}{ n_0} + O\left(\frac{1}{n_0^2}\right) ]
       \label{eq:ExpCeg}
    \end{aligned}
\end{equation}
where 
\begin{equation}
    \begin{aligned}
\delta C_{eg}\left(\theta,\Bar{n}\right) =  \frac{-2\theta + \left(3 + 2\bar{n}\right) \theta \cos \theta - \left(1 + 2\bar{n}\right)\left(1 + \theta^2\right) \sin \theta}{ 16} 
    \end{aligned}
\end{equation}

The expansions of $P_e(\theta)$ and $C_{eg}(\theta)$ can be used to find an expansion for the von Neumann entropy of the qubit up to first order in $n_0^{-1}$ in a similar way as
\begin{equation}
    \begin{aligned}
       S_\text{Q}(t) &= \epsilon_\text{Q}(\theta, \bar{n}) \left( - \frac{ \log\left(\epsilon_\text{Q}(\theta, \bar{n})/e\right)}{n_0} + \frac{\log n_0}{n_0} \right),
    \end{aligned}
    \label{eq:SQt}
\end{equation}
where
\begin{equation}
    \begin{aligned}
       \epsilon_\text{Q}(\theta, \bar{n}) &= \frac{(1 + 2\bar{n})(1 + 2\theta^2 - \cos2\theta ) + 4\theta \sin\theta}{32}.
    \end{aligned}
\end{equation}

\red{The term $\log n_0/{n_0}$ satisfies $ {1}/{n_0} \le \log n_0/{n_0} \le 1/\sqrt{n_0} $, for $n_0 \ge e$.}

 We use the same methodology to calculate the expansion of the thermodynamic quantities of the cavity. To determine those thermodynamics quantities, we observe from the numerics that, in the unitary regime, the cavity remains approximately in a Gaussian state which is thus fully characterized by $\langle a\rangle$, $\langle a^\dagger \rangle$, $\langle a^\dagger a \rangle$ and $\langle a^2\rangle$. 
 
Starting with the exact reduced state of the cavity for $\phi_0=0$
\begin{equation}
    \begin{aligned}
       \hat{\rho}_\text{C}\left(t\right) &=  \cos\left(g t \sqrt{a^{\dagger}a+1}\right) D\left(\alpha_0\right)\hat{w}_{\beta_C\left(0\right)}D^{\dagger}\left(\alpha_0\right) \cos\left(g t \sqrt{a^{\dagger}a+1}\right) +  \\
       & \frac{\sin\left({g t \sqrt{a^{\dagger}a }}\right)}{\sqrt{a^{\dagger}a }} a^{\dagger} D\left(\alpha_0\right)\hat{w}_{\beta_C\left(0\right)}D^{\dagger}\left(\alpha_0\right) a \frac{\sin\left({g t \sqrt{a^{\dagger}a }}\right)}{\sqrt{a^{\dagger}a }} 
       \label{eq:rhoC} 
    \end{aligned}
\end{equation}
the average $\langle a \rangle$ is given by 
\begin{equation}
    \begin{aligned}
       \langle a \rangle &= Tr[\hat{\rho}_\text{C}\left(t\right) a] \\
       &= \sum_n  [\cos\left(g t \sqrt{n+1}\right) \cos\left(g t \sqrt{n}\right)  +  {\sin\left({g t \sqrt{n +1 }}\right)} \sin\left({g t \sqrt{n}}\right) \frac{\sqrt{n+1}}{\sqrt{n}} ]\bra{n}  D\left(\alpha_0\right)\hat{w}_{\beta_C\left(0\right)}D^{\dagger}\left(\alpha_0\right) a\ket{n}  
      \end{aligned}
\end{equation}
which can be decomposed into     
\begin{equation}
    \begin{aligned}
  \langle a \rangle       &=  \sum_n \left(\Lambda_0 + \Lambda_1 n + \Lambda_2 n^2 \right) \bra{n} D\left(\alpha_0\right)\hat{w}_{\beta_C\left(0\right)}D^{\dagger}\left(\alpha_0\right) a \ket{n} + O\left( \frac{1}{n_0^4} \right) 
    \end{aligned}
\end{equation}
with 
\begin{equation}
\small
    \begin{aligned}
        \Lambda_0 = \frac{{1 + \frac{\theta^2}{4} + 2 n_0 \left(-3 + 4 n_0 \left(3 + 4 n_0\right) - \theta^2\right) + \left(-1 + 6 n_0 - 24 n_0^2 + \left(1 - 2 n_0\right)^2 \frac{\theta^2}{4}\right) \cos\left(\theta\right) -  \left(1 - 3 n_0\right)^2 \theta \sin\left(\theta\right)}}{{32 n_0^3}} 
    \end{aligned}
\end{equation}

\begin{equation}
\small
    \begin{aligned}
       \Lambda_1 = \frac{{1 - 6 n_0 + \frac{\theta^2}{4} + \left(-1 + \frac{\theta^2}{4} - 2 n_0 \left(-3 + \frac{\theta^2}{4}\right)\right) \cos\left(\theta\right) + \left(-2 + 7 n_0\right) \frac{\theta}{2} \sin\left(\theta\right)}}{{8 n_0^3}} \nonumber
    \end{aligned}
\end{equation}
\begin{equation}
\small
    \begin{aligned}
       \Lambda_2 = \frac{{4 + 2 \left(  \frac{\theta^2}{4}-2\right) \cos\left(\theta\right) -  \frac{5\theta}{2} \sin\left(\theta\right)}}{{16 n_0^3}} \nonumber
    \end{aligned}
\end{equation}
Summing up these terms we get
\begin{equation}
\small
    \begin{aligned}
        \langle a \rangle &= \sqrt{n_0} \Big(1+\frac{\sin^2{\frac{\theta}{2}}}{2 {n_0}}+\frac{ -2 \sin^2{\frac{\theta}{2}} \left(1 + \frac{\theta^2}{2} \right) + \frac{\theta}{2} \sin{\theta} +  \bar{n} \theta \left(\theta \cos{\theta} - \sin{\theta} \right) }{16 {n_0}^2} + \\
        & \frac{ 13 + \frac{5 \theta^2}{4} + 8 \Bar{n} \left(7 + 6 \Bar{n} + \frac{\theta^2}{4}\right) + \left(-13 + \frac{9 \theta^2}{4} + 8 \Bar{n} \left(-7 - 6 \bar{n} + \left(4 + 3 \Bar{n}\right) \frac{\theta^2}{4}\right)\right) \cos{\theta} - 2 \left(5 + \Bar{n} \left(19 + 15 \Bar{n}\right)\right) \theta \sin{\theta}
 }{32 n_0^3}\\
 & + O\left(\frac{1}{n_0^4}\right)  \Big) \; . 
 \label{eq:avga}
    \end{aligned}
\end{equation}
Similarly, we find
\begin{equation}
\small
    \begin{aligned}
        \langle a^2 \rangle &= n_0 \Big(1+\frac{\sin ^2{\frac{\theta}{2}}}{ {n_0}}+\frac{ \left(1 + 2\Bar{n}\right)[\left(1+\frac{\theta^2}{4}\right) \cos{\theta} - 1] + \left(3 + 2\bar{n}\right)\frac{\theta}{4} \sin{\theta} - \frac{\theta^2}{2} }{4 {n_0}^2} \\
        & + \frac{ 5 + \frac{3 \theta^2}{4} + 6\bar{n} \left(4 + 4 \bar{n} + \frac{\theta^2}{4}\right) + \left(-5 +  \theta^2 + 3 \bar{n} \left(-8 \left(1 + \bar{n}\right) + \left(5 + 4 \bar{n}\right) \frac{\theta^2}{4}\right)\right) \cos\theta - \left(17 + 69 \bar{n} + 60 \bar{n}^2\right) \frac{\theta}{4}  \sin{\theta}
 }{4 n_0^3}\\
 & + O\left(\frac{1}{n_0^4}\right) \Big) \; . 
 \label{eq:avga2}
    \end{aligned}
\end{equation}
Owing to the resonance condition, the total Hamiltonian $\hat H$ conserves the total number of excitations. Consequently, we have at any time:
\begin{equation}
    \begin{aligned}
       \langle a^\dagger a  \rangle &= \bar{n} + n_0 + 1 - P_e\left(t\right) \label{eq:avgadafull}
    \end{aligned}
\end{equation}
Using the expansion of $P_e\left(t\right)$ up to second order Eq.~\eqref{eq:ExpPe}, we simplify the above term as
\begin{equation}
    \begin{aligned}
         \langle a^\dagger a\rangle &= \bar{n} + n_0 + \sin^2{\frac{\theta}{2}} + \frac{\theta\left( \left(1+2\Bar{n}\right) \theta \cos{\theta} + \left(3+2\Bar{n} \right)\sin{\theta} \right)}{16 n_0} + \frac{\theta \left(1+\Bar{n}\right)\left(1 + 2\Bar{n}\right)\left(\theta \cos{\theta} - \sin{\theta}\right)}{16n_0^2}\\
         & + O\left(\frac{1}{n_0^3}\right) \; . 
         \label{eq:avgada}
    \end{aligned}
\end{equation}

The covariance matrix is defined as:
\begin{equation}
    \begin{aligned}
       V &= \begin{bmatrix}
\sigma_q^2 & \sigma_{qp} \\
\sigma_{pq} & \sigma_p^2 \\
\end{bmatrix}
    \end{aligned} \label{eq:variancemat}
\end{equation}
with
$$ \sigma_{xy} = \frac{1}{2} \langle \hat{x}\hat{y}  + \hat{y} \hat{x}\rangle- \langle \hat{x} \rangle \langle \hat{y} \rangle, \hspace{5pt} \left(x,y\right) \in \{q,p\},\hspace{1pt} \hat{q} = \frac{a + a^\dagger}{\sqrt{2}}, \hspace{1pt} \hat{p} = \frac{a - a^\dagger}{i \sqrt{2}} $$

Since $\phi_0 = 0$, the covariance matrix $V$ takes the diagonal form 
\begin{equation}
    \begin{aligned}
       V &= \begin{bmatrix}
\langle a^2 \rangle + \langle a^\dagger a \rangle + \frac{1}{2}- 2 \langle a \rangle^2 & 0 \\
0 & -\langle a^2 \rangle + \langle a^\dagger a \rangle + \frac{1}{2} \\
\end{bmatrix} \; . 
    \end{aligned} \nonumber
\end{equation}

For Gaussian states, the von Neumann entropy can be evaluated as \cite{agarwal1971entropy}
\begin{equation}
    \begin{aligned}
       S_\text{C}\left(t\right) &= h\left( \sqrt{Det[V]} \right), \hspace{5pt} h\left(x\right) = \left(x+\frac{1}{2}\right) \log\left({x + \frac{1}{2}}\right) - \left(x-\frac{1}{2}\right) \log\left({x - \frac{1}{2}}\right) \; . 
    \end{aligned}
\end{equation}
Using the previous expansions the determinant is expressed as
\begin{equation}
    \begin{aligned}
        \sqrt{Det[V]} & = \left(\Bar{n}+\frac{1}{2}\right) \sqrt{[1 + \frac{1}{4 n_0 \left(2 \Bar{n} + 1\right)} \left( \theta^2 + \sin^2{\theta} + 2 \left(1+2\Bar{n}\right)\theta \sin{\theta} \right) ]} \label{eq:detV}
    \end{aligned}
\end{equation}

Using the framework  of \cite{Elouard23} we can calculate the heat exchanges of the cavity. We must first find the thermal state $\hat{\rho}_\text{C}^{th}\left(t\right)$ with the same von Neumann entropy $S_\text{C}\left(t\right)$ to do so. The von Neumann entropy of a thermal state with photon numbers $n_\text{C}$ is given by $ \log[\frac{n_\text{C}^{n_\text{C}}}{\left(n_\text{C}+1\right)^{n_\text{C}+1}}]$. Comparing the entropies    

$$ \red{-\log[\frac{n_\text{C}^{n_\text{C}}}{\left(n_\text{C}+1\right)^{n_\text{C}+1}}] = S_\text{C}\left(t\right)  \implies n_\text{C} + \frac{1}{2} = \sqrt{Det[V]}}$$

To find the heat exchanges $Q_\text{C}$, we recall that the heat is defined as the negative of the variation in the thermal energy of the cavity, which takes the following form
$ Q_\text{C}\left(t\right)  =  - Tr[\hat{H}_\text{C} \left(\hat{\rho}_\text{C}^{th}\left(t\right) - \hat{\rho}_\text{C}^{th}\left(0\right)\right)] $ 
and reads
\begin{equation}
    \begin{aligned}
       Q_\text{C}\left(t\right) = \hbar \omega_0 \left(\Bar{n}+\frac{1}{2}\right) \left[1 -  \sqrt{1 + \frac{1}{4 n_0 \left(2 \Bar{n} + 1\right)} \left( \theta^2 + \sin^2{\theta} + 2 \left(1+2\Bar{n}\right)\theta \sin{\theta} \right) }\right]  \;
       \label{eq:QC}
    \end{aligned}
\end{equation}
which up to first order in $1/n_0$ yields:
\begin{equation}
    \begin{aligned}
       Q_\text{C}\left(t\right) = -   \frac{\hbar \omega_0}{16 n_0} \left( \theta^2 + \sin^2{\theta} + 2 \left(1+2\Bar{n}\right)\theta \sin{\theta} \right) + O(\frac{1}{n_0^2}) \; . 
    \end{aligned}
    \label{eq:QC1}
\end{equation}
By using the expression for $\langle a^\dagger a\rangle$, we can also determine the change in the cavity's internal energy as follows
\begin{equation}
    \begin{aligned}
      \Delta E_\text{C}\left(t\right) = \hbar \omega_0 \left(1 - P_e\left(t\right)\right)
       \label{eq:DEC}
    \end{aligned}
\end{equation}
In the end, the work done by the cavity is given by
\begin{equation}
    \begin{aligned}
     W_\text{C}\left(t\right) = - \Delta E_\text{C}\left(t\right) - Q_\text{C}\left(t\right)\; . 
       \label{eq:WC}
    \end{aligned}
\end{equation}

Using Eq.~\eqref{eq:SQt} and Eq.~\eqref{eq:QC1}, we can write the entropy production denoted as $\sigma_\text{Q}$ of the qubit system as

\begin{equation}
    \begin{aligned}
       \sigma_\text{Q} = S_\text{Q}(t) - \beta_C(0) Q_\text{C}(t) \ge 0
    \end{aligned}
    \label{eq:sigmaQ}
\end{equation}

\section{Joint atom-cavity evolution on longer times}

\subsection{Perturbative expansion of the global atom–cavity state}\label{s:Meas}
In this section, we provide the expressions for cavity and \red{atom evolution} over the interval $[0,t_c]$, when the system starts in state:

\bb
\hat{\rho}\left(0\right) = \sum_{\nu,\mu = \pm} \rho_{\nu,\mu} \ket{\nu_y}\bra{\mu_y} \otimes \hat{\rho}_\text{C}\left(0\right), \hspace{5pt} \hat{\rho}_\text{C}\left(0\right) = D\left(\alpha_0\right) \hat{w}_{\beta_C\left(0\right)} D^\dagger\left(\alpha_0\right),\hspace{5pt} \alpha_0 = \sqrt{n_0}
\ee
where in $\{\ket{e},\ket{g}\}$ basis:
\bb
\ket{\pm_y}\bra{\pm_y} = \frac{1}{2} \begin{bmatrix}
1  & \mp i  \\
\pm i  & 1 
\end{bmatrix},\hspace{10pt} \ket{\pm_y}\bra{\mp_y} = \frac{1}{2} \begin{bmatrix}
1  & \pm i  \\
\pm i  & -1 
\end{bmatrix}.
\ee

It is useful to recast the total evolution operator as
\begin{equation}
    \hat{U}\left(t\right) = \hat{c}_1 \ket{e}\bra{e} + \hat{c}_0 \ket{g}\bra{g}  + \hat{s}_1 \ket{e}\bra{g} - \hat{s}_0 \ket{g}\bra{e}, 
\end{equation}
with $ \hat{c}_1= \cos\left(gt\sqrt{a^\dagger a + 1}\right),\hspace{1pt} \hat{c}_0= \cos\left(gt\sqrt{a^\dagger a}\right), \hspace{1pt} \hat{s}_1 = \frac{\sin\left(gt \sqrt{a^\dagger a  + 1}\right)}{\sqrt{a^\dagger a +1}} a, \hspace{1pt} \hat{s}_0 = \frac{\sin\left(gt \sqrt{a^\dagger a  }\right)}{\sqrt{a^\dagger a }} a^\dagger$ and to introduce
\bb 
\hat{s}_1^{'} = \frac{\sin\left(g t \sqrt{\hat{n} + 1}\right)}{\sqrt{\hat{n} + 1}},\hspace{5pt} \hat{s}_0^{'} = \frac{\sin\left(g t \sqrt{\hat{n}}\right)}{\sqrt{\hat{n}}} 
\ee

Below, we compute separately the evolution of the basis of qubit operators $\{\ket{\nu_y}\bra{\mu_y}\}$, $\mu,\nu=\pm$. 

\begin{equation}
              \begin{aligned}
                        \hat{U}\left(t\right) \ket{+_y}\bra{+_y} \otimes \hat{\rho}_\text{C}\left(0\right) \hat{U}^\dagger\left(t\right) &=  \hat{U}\left(t\right) \frac{1}{2}[ \ket{e}\bra{e} - i  \ket{e}\bra{g} + i  \ket{g}\bra{e} + \ket{g}\bra{g} ] \otimes \hat{\rho}_\text{C}\left(0\right) \hat{U}^\dagger\left(t\right) \\
                        &= \sum_{p,q=e,g} \ket{p}\bra{q} \otimes \bra{p} \hat{U}\left(t\right) \ket{+_y}\bra{+_y} \otimes \hat{\rho}_\text{C}\left(0\right) \hat{U}^\dagger\left(t\right) \ket{q} \\
                        &=  \sum_{p,q=e,g} \hat{T}_{pq},
                        \label{eq:PYPY1}
              \end{aligned}
          \end{equation}
    where $ \hat{T}_{pq} = \ket{p}\bra{q} \otimes \bra{p} \hat{U}\left(t\right) \ket{+_y}\bra{+_y} \otimes \hat{\rho}_\text{C}\left(0\right) \hat{U}^\dagger\left(t\right) \ket{q} $. For $p=e$ and $q=e$:
             \begin{equation}
                 \begin{aligned}
                    \hat{T}_{ee} &= \frac{1}{2} \ket{e}\bra{e} \otimes [ \hat{c}_1 \hat{\rho}_\text{C}\left(0\right) \hat{c}_1 + \hat{s}_1 \hat{\rho}_\text{C}\left(0\right) \hat{s}_1^\dagger - i  \hat{c}_1 \hat{\rho}_\text{C}\left(0\right) \hat{s}_1^\dagger + i \hat{s}_1 \hat{\rho}_\text{C}\left(0\right) \hat{c}_1 ]. \label{eq:Tee}
                 \end{aligned}
             \end{equation}
            
             We can use the definition of $\hat{s}^{'}_{0,1}$ along with the relations $D^\dagger\left(\alpha_0\right) a D\left(\alpha_0\right) = a + \alpha$ and $D^\dagger\left(\alpha_0\right) a^\dagger D\left(\alpha_0\right) = a^\dagger + \alpha^*$ to transform the above into the following
             
             \begin{equation}
     \begin{aligned}
     \hat{T}_{ee} &=  \frac{1}{2} \ket{e}\bra{e} \otimes [ \hat{c}_1 \hat{\rho}_\text{C}\left(0\right) \hat{c}_1 + \hat{s}_1^{'} a \hat{\rho}_\text{C}\left(0\right) a^\dagger \hat{s}_1^{'} - i  \hat{c}_1 \hat{\rho}_\text{C}\left(0\right) a^\dagger \hat{s}_1{'} + i \hat{s}_1^{'} a \hat{\rho}_\text{C}\left(0\right) \hat{c}_1 ] \\
      &=  \frac{1}{2} \ket{e}\bra{e} \otimes \Big[ \hat{c}_1 \hat{\rho}_\text{C}\left(0\right) \hat{c}_1 + \\
      &\hat{s}_1^{'} [ n_0 \hat{\rho}_\text{C}\left(0\right) + D\left(\alpha_0\right) \hat{w}_{\beta_C\left(0\right)} a^\dagger D^\dagger\left(\alpha_0\right) \alpha + D\left(\alpha_0\right) a \hat{w}_{\beta_C\left(0\right)}  D^\dagger\left(\alpha_0\right) \alpha^* + D\left(\alpha_0\right)a \hat{w}_{\beta_C\left(0\right)} a^\dagger D^\dagger\left(\alpha_0\right) ] \hat{s}_1^{'} \\
      & - i  \hat{c}_1 \left[\alpha^* \hat{\rho}_\text{C}\left(0\right) + D\left(\alpha_0\right) \hat{w}_{\beta_C\left(0\right)} a^\dagger D^\dagger\left(\alpha_0 \right) \right] \hat{s}_1{'} +  i \hat{s}_1^{'} [ \alpha \hat{\rho}_\text{C}\left(0\right) + D\left(\alpha_0\right) a \hat{w}_{\beta_C\left(0\right)} D^\dagger\left(\alpha_0\right) ] \hat{c}_1 \Big] \\
      \label{eq:Tee1}
     \end{aligned}
 \end{equation}
 The terms that include at least one bosonic operator acting on the thermal state, such as $ D\left(\alpha_0\right) \hat{w}_{\beta_C\left(0\right)} a^\dagger D^\dagger\left(\alpha_0\right) \alpha $, $ D\left(\alpha_0\right) a \hat{w}_{\beta_C\left(0\right)} D^\dagger\left(\alpha_0\right) \alpha^* $, and $ D\left(\alpha_0\right) a \hat{w}_{\beta_C\left(0\right)} a^\dagger D^\dagger\left(\alpha_0\right) $, have a vanishing magnitude when tracing over the cavity, and are therefore neglected, leading to:     

 \begin{equation}
     \begin{aligned}
     \hat{T}_{ee} &= \frac{1}{2} \ket{e}\bra{e} \otimes \left(\hat{c}_1 + i \alpha  \hat{s}_1^{'}\right) \hat{\rho}_\text{C}\left(0\right) \left(\hat{c}_1 - i \alpha^*  \hat{s}_1^{'}\right)  \\
       &= \frac{1}{2} \ket{e}\bra{e} \otimes \left(\hat{c}_1 + i \sqrt{n_0}  \hat{s}_1^{'}\right) \hat{\rho}_\text{C}\left(0\right) \left(\hat{c}_1 - i \sqrt{n_0} \hat{s}_1^{'}\right) \label{eq:Tee2}
     \end{aligned}
 \end{equation}
We can now use the following approximation, applicable for large $n_0$ to simplify the expression above:
\begin{equation}
    \begin{aligned}
     \hat{c}_1+ i \sqrt{n_0} \hat{s}_1^{'} &=  \cos\left(g t \sqrt{\hat{n}+1}\right) + i \sqrt{n_0}  \frac{\sin\left(g t \sqrt{\hat{n} + 1}\right)}{\sqrt{\hat{n} + 1}}   \\
     &= \cos\left(g t \sqrt{\hat{n} + 1}\right) + i  \sin\left(g t \sqrt{\hat{n} + 1}\right) [ 1 - \frac{1}{2} \frac{\hat{n} + 1 - n_0}{n_0}  + .. ] \\
    &= e^{ig t \sqrt{\hat{n} + 1}} + O\left(\frac{1}{n_0}\right)  \\
    &= e^{\frac{ig t \sqrt{n_0}}{2}} e^{ \frac{ig t \left(\hat{n} + 1\right) }{2\sqrt{n_0}} }  + O\left(\frac{1}{n_0}\right). 
    \label{eq:appr1}
    \end{aligned}
\end{equation}

We obtain:
 \begin{equation}
     \begin{aligned}
     \hat{T}_{ee} 
       &= \frac{1}{2} \ket{e}\bra{e} \otimes e^{\frac{ig t \hat{n}}{2\sqrt{n_0}}} \hat{\rho}_\text{C}\left(0\right) e^{-\frac{ig t \hat{n}}{2\sqrt{n_0}}} + O\left(\frac{1}{n_0}\right). 
       \label{eq:Teefinal}
     \end{aligned}
 \end{equation}
 
We proceed similarly to simplify the other terms $\hat T_{pq}$. For $p = e $ and $ q = g $:
             
\begin{equation}
     \begin{aligned}
     \hat{T}_{eg} =\hat{T}_{ge}^\dagger &= \frac{1}{2} \ket{e}\bra{g} \otimes [ -\hat{c}_1 \hat{\rho}_\text{C}\left(0\right) \hat{s}_0^\dagger + \hat{s}_1 \hat{\rho}_\text{C}\left(0\right) \hat{c}_0 - i  \hat{c}_1 \hat{\rho}_\text{C}\left(0\right) \hat{c}_0 - i  \hat{s}_1 \hat{\rho}_\text{C}\left(0\right) \hat{s}_0^\dagger ] \\
      &= \frac{-i }{2} \ket{e}\bra{g} \otimes \left(\hat{c}_1 + i  \hat{s}_1\right) \hat{\rho}_\text{C}\left(0\right) \left(\hat{c}_0 - i \hat{s}_0^\dagger\right)\\
       &= \frac{-i }{2} \ket{e}\bra{g} \otimes \left(\hat{c}_1 + i \sqrt{n_0} \hat{s}_1^{'}\right) \hat{\rho}_\text{C}\left(0\right) \left(\hat{c}_0 - i \sqrt{n_0} \hat{s}_0^{'}\right)\nonumber\\
       &=-\frac{i e^{\frac{ig t }{2 \sqrt{n_0}}}}{2} \ket{e}\bra{g} \otimes e^{\frac{ig t \hat{n}}{2\sqrt{n_0}}} \hat{\rho}_\text{C}\left(0\right) e^{-\frac{ig t \hat{n}}{2\sqrt{n_0}}} + O\left(\frac{1}{n_0}\right)  \label{eq:Teg}
     \end{aligned}
 \end{equation}
We approximated term $\hat{c}_0 - i \sqrt{n_0} \hat{s}_0^{'}$ using:

\begin{equation}
    \begin{aligned}
     \hat{c}_0+ i \sqrt{n_0} \hat{s}_0^{'} &=  \cos\left(g t \sqrt{\hat{n}}\right) + i \sqrt{n_0}  \frac{\sin\left(g t \sqrt{\hat{n} }\right)}{\sqrt{\hat{n} }}   \\
     &= \cos\left(g t \sqrt{\hat{n}}\right) + i  \sin\left(g t \sqrt{\hat{n} }\right) [ 1 - \frac{1}{2} \frac{\hat{n} - n_0}{n_0}  + .. ] \\
    &= e^{ig t \sqrt{\hat{n}}} + O\left(\frac{1}{n_0}\right)  \\
    &= e^{\frac{ig t \sqrt{n_0}}{2}} e^{ \frac{ig t \hat{n} }{2\sqrt{n_0}} }  + O\left(\frac{1}{n_0}\right) 
    \label{eq:appr2}
    \end{aligned}
\end{equation}

  Finally, for $p=g$ and $q=g$
          
 \begin{equation}
     \begin{aligned}
      \hat{T}_{gg}
       &= \frac{1}{2} \ket{g}\bra{g} \otimes e^{\frac{ig t \hat{n}}{2\sqrt{n_0}}} \hat{\rho}_\text{C}\left(0\right) e^{-\frac{ig t \hat{n}}{2\sqrt{n_0}}} + {\cal O}\left(\frac{1}{n_0}\right)   
       \label{eq:Tgg}
     \end{aligned}
 \end{equation}
             
 
Collecting the four contributions:
\begin{equation}
    \begin{aligned}
      \hat{U}\left(t\right) \ket{+_y}\bra{+_y} \otimes \hat{\rho}_\text{C}\left(0\right) \hat{U}^\dagger\left(t\right) &= \frac{1}{2}\left[ \begin{matrix}  1 & -i e^{\frac{ig t}{2\sqrt{n_0}}}\\
      i e^{-\frac{ig t}{2\sqrt{n_0}}} & 1 \end{matrix} \right] \otimes D\left(\alpha_0 e^{\frac{igt}{2\sqrt{n_0}}}\right)\hat{w}_{\beta_C\left(0\right)} D^\dagger\left(\alpha_0 e^{\frac{igt}{2\sqrt{n_0}}}\right)+ {\cal O}\left(\frac{1}{n_0}\right) 
      \label{eq:pypyevol}
    \end{aligned}
\end{equation}

Using the same method, we compute the evolution of the three other basis elements:

\begin{equation}
    \begin{aligned}
      \hat{U}\left(t\right) \ket{-_y}\bra{-_y} \otimes \hat{\rho}_\text{C}\left(0\right) \hat{U}^\dagger\left(t\right) &= \frac{1}{2}\left[ \begin{matrix}  1 & i e^{-\frac{ig t}{2\sqrt{n_0}}}\\
      -i e^{\frac{ig t}{2\sqrt{n_0}}} & 1 \end{matrix} \right] \otimes D\left(\alpha_0 e^{-\frac{igt}{2\sqrt{n_0}}}\right)\hat{w}_{\beta_C\left(0\right)} D^\dagger\left(\alpha_0 e^{-\frac{igt}{2\sqrt{n_0}}}\right) 
      \label{eq:mymyevol}
    \end{aligned}
\end{equation}

\begin{equation}
    \begin{aligned}
      \hat{U}\left(t\right) \ket{+_y}\bra{-_y} \otimes \hat{\rho}_\text{C}\left(0\right) \hat{U}^\dagger\left(t\right) &= \frac{e^{i g t \sqrt{n_0}}}{2}\left[ \begin{matrix}  e^{\frac{ig t}{\sqrt{n_0}}} & i e^{\frac{ig t}{2\sqrt{n_0}}}\\
      i e^{\frac{ig t}{2\sqrt{n_0}}} & -1 \end{matrix} \right] \otimes D\left(\alpha_0 e^{\frac{igt}{2\sqrt{n_0}}}\right)\hat{w}_{\beta_C\left(0\right)} D^\dagger
      \left(\alpha_0 e^{-\frac{igt}{2\sqrt{n_0}}}\right) \label{eq:pymyevol}
    \end{aligned}
\end{equation}

\begin{equation}
    \begin{aligned}
      \hat{U}\left(t\right) \ket{-_y}\bra{+_y} \otimes \hat{\rho}_\text{C}\left(0\right) \hat{U}^\dagger\left(t\right) &= \frac{e^{-i g t \sqrt{n_0}}}{2}\left[ \begin{matrix}  e^{-\frac{ig t}{\sqrt{n_0}}} & -i e^{-\frac{ig t}{2\sqrt{n_0}}}\\
      -i e^{\frac{-ig t}{2\sqrt{n_0}}} & -1 \end{matrix} \right] \otimes D\left(\alpha_0 e^{-\frac{igt}{2\sqrt{n_0}}}\right)\hat{w}_{\beta_C\left(0\right)} D^\dagger
      \left(\alpha_0 e^{\frac{igt}{2\sqrt{n_0}}}\right) \label{eq:mypyevol}
    \end{aligned}
\end{equation}

We deduce the evolution of the system starting from an arbitrary initial qubit state:

\begin{eqnarray}
    \hat{\rho}_\text{QC}\left(t\right) &=& \hat{U}\left(t\right) \sum_{\nu,\mu = \pm} \rho_{\nu,\mu} \ket{\nu_y}\bra{\mu_y}  \otimes D\left(\sqrt{n_0}\right) \hat{w}_{\beta_C\left(0\right)} D^\dagger\left(\sqrt{n_0}\right) \hat{U}^\dagger\left(t\right)   
    \label{eq:rhoQCt0}\nonumber\\
    &=& \frac{\rho_{++}}{2}\left[ \begin{matrix}  1 & -i e^{\frac{ig t}{2\sqrt{n_0}}}\\
      i e^{-\frac{ig t}{2\sqrt{n_0}}} & 1 \end{matrix} \right] \otimes D\left(\alpha_0 e^{\frac{igt}{2\sqrt{n_0}}}\right)\hat{w}_{\beta_C\left(0\right)} D^\dagger\left(\alpha_0 e^{\frac{igt}{2\sqrt{n_0}}}\right) \nonumber\\
      && + \frac{\rho_{--}}{2}\left[ \begin{matrix}  1 & i e^{\frac{-ig t}{2\sqrt{n_0}}}\\
      -i e^{\frac{ig t}{2\sqrt{n_0}}} && 1 \end{matrix} \right] \otimes D\left(\alpha_0 e^{-\frac{igt}{2\sqrt{n_0}}}\right)\hat{w}_{\beta_C\left(0\right)} D^\dagger\left(\alpha_0 e^{-\frac{igt}{2\sqrt{n_0}}}\right) +\nonumber \\
      &&+ \frac{ \rho_{+-}e^{i g t \sqrt{n_0}}}{2}\left[ \begin{matrix}  e^{\frac{ig t}{\sqrt{n_0}}} & i e^{\frac{ig t}{2\sqrt{n_0}}}\nonumber\\
      i e^{\frac{ig t}{2\sqrt{n_0}}} && -1 \end{matrix} \right] \otimes D\left(\alpha_0 e^{\frac{igt}{2\sqrt{n_0}}}\right)\hat{w}_{\beta_C\left(0\right)} D^\dagger
      \left(\alpha_0 e^{-\frac{igt}{2\sqrt{n_0}}}\right) \nonumber\\
      && + \frac{\rho_{-+} e^{-i g t \sqrt{n_0}}}{2}\left[ \begin{matrix}  e^{-\frac{ig t}{\sqrt{n_0}}} & -i e^{-\frac{ig t}{2\sqrt{n_0}}}\nonumber\\
      -i e^{\frac{-ig t}{2\sqrt{n_0}}} & -1 \end{matrix} \right] \otimes D\left(\alpha_0 e^{-\frac{igt}{2\sqrt{n_0}}}\right)\hat{w}_{\beta_C\left(0\right)} D^\dagger
      \left(\alpha_0 e^{\frac{igt}{2\sqrt{n_0}}}\right) + O\left(\frac{1}{n_0}\right) 
      \label{eq:rhoQCt1}\\
   &=&  \rho_{++} \ket{+_y\left(t\right)}\bra{+_y\left(t\right)} \otimes \hat{\rho}_\text{C}^{++}\left(t\right)+ \rho_{--} \ket{-_y\left(t\right)}\bra{-_y\left(t\right)} \otimes \hat{\rho}_\text{C}^{--}\left(t\right)\nonumber \\
   &&+ \rho_{+-} e^{ig t \sqrt{n_0} + \frac{i g t }{\sqrt{n_0}}} \ket{+_y\left(t\right)}\bra{-_y\left(t\right)} \otimes \hat{\rho}_\text{C}^{+-}\left(t\right)+ + \rho_{-+} e^{-ig t \sqrt{n_0} - \frac{i g t }{\sqrt{n_0}}} \ket{-_y\left(t\right)}\bra{+_y\left(t\right)} \otimes \hat{\rho}_\text{C}^{-+}\left(t\right)\nonumber\\
   &&+ O\left(\frac{1}{n_0}\right),
   \label{eq:rhoQCt2}
\end{eqnarray}
where we have introduced the states:
\bb \label{pmyt}
\ket{\pm_y\left(t\right)} = \frac{1}{\sqrt{2}} \left(\begin{matrix} 1 \\ \pm i e^{\mp \frac{igt}{2\sqrt{n_0}}} \end{matrix}\right),
\ee
and $ \hat{\rho}_\text{C}^{\nu\mu}\left(t\right)$ is defined in Eq.~\eqref{eq:rhoCnumu} of main text.

We now show that the trace of the terms $\hat{\rho}_\text{C}^{\pm \mp}\left(t_c\right)$ becomes negligible at $t_c$. We have:
\begin{equation}
    \begin{aligned}
        \text{Tr}[\hat{\rho}_\text{C}^{\pm \mp}\left(t\right)] &=e^{\pm igt(\sqrt{n_0} + 1/\sqrt{n_0})} \text{Tr}[D\left(\sqrt{n_0} e^{\pm i\frac{gt}{2\sqrt{n_0}}}\right) \hat{w}_{\beta_C\left(0\right)} D^\dagger\left(\sqrt{n_0} e^{\mp i\frac{gt}{2\sqrt{n_0}}}\right)] \\
        &= e^{\pm igt(\sqrt{n_0} + 1/\sqrt{n_0})} e^{ \pm i n_0 \sin\frac{gt}{\sqrt{n_0}} } \text{Tr}[ \hat{w}_{\beta_C\left(0\right)} D\left(\pm 2i\sqrt{n_0} \sin\left(\frac{gt}{2\sqrt{n_0}}\right) \right) ] \\
        &= e^{\pm igt(\sqrt{n_0} + 1/\sqrt{n_0})} \left(1-e^{-\beta_C\left(0\right)}\right) e^{ \pm i n_0 \sin{\frac{gt}{\sqrt{n_0}}} } \frac{1}{\pi} \int d^{\left(2\right)}\alpha  \bra{\alpha} e^{-\beta_C\left(0\right) a^\dagger a} D\left(\pm 2i\sqrt{n_0} \sin\left(\frac{gt}{2\sqrt{n_0}}\right)\right) \ket{\alpha} 
         \nonumber
    \end{aligned}
\end{equation}

On using the known properties of coherent states and displacement operator to simplify the integrand above and carrying out the Gaussian integrals, we obtain

\begin{equation}
    \begin{aligned}
       \text{Tr}[\hat{\rho}_\text{C}^{\pm \mp}\left(t\right)] &= e^{\pm igt(\sqrt{n_0} + 1/\sqrt{n_0})}e^{ \pm i n_0 \sin{\frac{gt}{\sqrt{n_0}}} } e^{ -2n_0 \sin^2{\frac{gt}{2\sqrt{n_0}}} } e^{-\frac{4n_0 \bar{n}}{\left(1 + \bar{n}\right)^2}\sin^2{\frac{gt}{2\sqrt{n_0}}} } \label{eq:trrhopm0}
    \end{aligned}
\end{equation}
 As $n_0 \to \infty$
 
 \begin{equation}
    \begin{aligned}
       \text{Tr}[\hat{\rho}_\text{C}^{\pm \mp}\left(t\right)] &= e^{ \pm 2i gt \sqrt{n_0}  }e^{ -\frac{g^2 t^2}{2} } e^{-\frac{g^2 t^2 \bar{n}}{\left(1 + \bar{n}\right)^2} + \frac{\bar{n}}{\left(1+\bar{n}\right)^2} \frac{g^4 t^4}{12 n_0} }, \label{eq:trrhopm1}
    \end{aligned}
\end{equation}
 which tends to $0$.
 
Consequently, we can express the reduced state of the atom at time $t_c$ as:

\begin{equation}
    \begin{aligned}
   \hat{\rho}_\text{Q}\left(t_c\right) &=  \rho_{++} \ket{+_y\left(t_c\right)}\bra{+_y\left(t_c\right)}+ \rho_{--} \ket{-_y\left(t_c\right)}\bra{-_y\left(t_c\right)}  + O\left(\frac{1}{n_0}\right) 
   \label{eq:rhoQtc}
    \end{aligned}
\end{equation}

We can see that for times $t\in [0,g^{-1}]$, the term $e^{\pm \frac{ig t}{2\sqrt{n_0}}}$ approaches 1 as $n_0$ approaches infinity, which allows us to rewrite the state of the qubit at $t_c$ as:

\begin{equation}
    \begin{aligned}
   \hat{\rho}_\text{Q}\left(t_c\right) &=  \rho_{++} \ket{+_y}\bra{+_y}+ \rho_{--} \ket{-_y}\bra{-_y}  + O\left(\frac{1}{n_0}\right) 
   \label{eq:rhoQtc1}
    \end{aligned}
\end{equation}

\subsection{Autonomous feedback}\label{s:Feedback}

On a longer timescale of order $t \sim \sqrt{n_0} \pi g^{-1}$, the cavity exerts an effective drive on the qubit that is dependent on the measurement result. Here we analyze the total dynamics along two branches associated with the qubit found in $\ket{\pm_y}$. As we are in the interaction picture, the total system evolves as follows:

\begin{equation}
    \begin{aligned}
       i\hbar \frac{d \hat{\rho}_\text{QC}(t) }{dt} &= \left[\hat{V}_\text{QC},  \hat{\rho}_\text{QC}(t) \right].
       \label{eq:timeevol}
    \end{aligned}
\end{equation}


Since the qubit-cavity state is factorized within each branch up to an order of $O\left(\frac{1}{n_0}\right)$, we can trace over the cavity space and get the effective evolution for qubit. Denoting $\hat\rho_\text{Q}^{\pm\pm}(t)$ the qubit states along each branch, with $\hat\rho_\text{Q}^{\pm\pm}(0)=\ket{\pm_y}\bra{\pm_y}$, we have:

\begin{equation}
    \begin{aligned}
       i\hbar \frac{d}{dt} \hat\rho_\text{Q}^{\pm\pm}(t) &= \left[i\hbar g ( \alpha_{\pm}(t)  \sigma_+ -  \alpha_{\pm}^*(t) \sigma_- ),  \hat\rho_\text{Q}^{\pm\pm}(t) \right] 
       \label{eq:timeevol1}
    \end{aligned}
\end{equation}

From Eq.~\eqref{eq:timeevol1}, we obtain the effective Hamiltonian acting on qubit

\begin{equation}
    \begin{aligned}
       \hat{H}_\text{Q}^{\text{eff}(\nu)}(t) &= i\hbar g( \alpha_{\nu}(t)  \sigma_+ -  \alpha_{\nu}^*(t) \sigma_- )
       \label{eq:HeffQ}.
    \end{aligned}
\end{equation}
\subsection{\red{Thermodynamic analysis of the Maxwell demon}}\label{s:ThermoDemon}

Using the perturbative expression Eq.~\eqref{eq:rhoQCt2} for the joint cavity-qubit state, we have

\begin{equation}
    \begin{aligned}
    \hat{\rho}_{QC}(t) = \sum_{\nu,\mu=\pm}\rho_{\nu\mu}  \ket{\nu_y(t)}\bra{\mu_y(t)} \otimes \hat{\rho}_\text{C}^{\nu\mu}(t) + {\cal O}\left(\frac{1}{n_0}\right),\label{eq:c17perturbative}
    \end{aligned}
\end{equation}

where $\rho_{\nu\mu} =  \bra{\nu_y}\hat{\rho}_\text{Q}(0)\ket{\mu_y}$ are the coefficients of the qubit density operator in the measurement basis, the states $\ket{\pm_y(t)}$ are defined in Eq.~\eqref{pmyt} and
\bb
\hat{\rho}_\text{C}^{\nu\mu}(t) \!\!\!&=&\!\!\! e^{\nu i \chi \delta_{\nu\mu}} D(\sqrt{n_0} e^{\nu \frac{igt}{2\sqrt{n_0}}}) \hat{w}_{\beta_C(0)} D^\dagger(\sqrt{n_0} e^{\mu \frac{igt}{2\sqrt{n_0}}}),\;\;\;\;
\ee
are conditional operators in the cavity space with $\chi = gt(\sqrt{n_0} + 1/\sqrt{n_0})$ and $\delta_{\nu\mu}$ the Kronecker delta. 

The global state consists of cavity and the qubit, under the condition that the qubit starts from maximally mixed state, takes the following form:

\begin{equation}
    \begin{aligned}
    \hat{\rho}_{QC}(t) = \frac{1}{2}  \ket{+_y(t)}\bra{+_y(t)} \otimes \hat{\rho}_\text{C}^{++}(t)+ \frac{1}{2}  \ket{-_y(t)}\bra{-_y(t)} \otimes \hat{\rho}_\text{C}^{--}(t) + {\cal O}\left(\frac{1}{n_0}\right) \label{eq:rhoQCM}.
    \end{aligned}
\end{equation}

\textbf{{\it Entropy of the Cavity.}} The reduced state of the cavity can be deduced from  Eq.~\eqref{eq:rhoQCM} as

\begin{equation}
    \begin{aligned}
        \hat{\rho}_\text{C}\left(t\right) &=
  \frac{1}{2} \hat{\rho}_\text{C}^{++}\left(t\right)+ \frac{1}{2}  \hat{\rho}_\text{C}^{--}\left(t\right) 
   + O\left(\frac{1}{n_0}\right), 
   \label{eq:rhoCavity2}
    \end{aligned}
\end{equation}

The states of the cavity $\hat{\rho}_C^{++}$, and $\hat{\rho}_C^{--}$ emerging in the two different branches have a finite overlap ($\text{Tr}[\hat{\rho}_C^{++} \hat{\rho}_C^{--}] \neq 0 $) during the measurement phase. As soon as the measurement ends on a timescale of $gt \sim 1$, these two states become orthogonal and macroscopically distinct. The overlap between $\text{Tr}[\hat{\rho}_C^{++} \hat{\rho}_C^{--}]$ can be calculated by evaluating the integration below, and it has the following form:

\begin{equation}
    \begin{aligned}
      \text{Tr}[\hat{\rho}_C^{++} \hat{\rho}_C^{--}] &= \text{Tr}[ D(2i\sqrt{n_0} \sin(\frac{gt}{2\sqrt{n_0}})) \hat{w}_{\beta_C(0)}D^\dagger(2i\sqrt{n_0} \sin(\frac{gt}{2\sqrt{n_0}})) \hat{w}_{\beta_C(0)} ] +  O\left(\frac{1}{n_0}\right)\\
      &= \frac{1}{\pi} \int d^2 \alpha \bra{\alpha} D(2i\sqrt{n_0} \sin(\frac{gt}{2\sqrt{n_0}})) \hat{w}_{\beta_C(0)}D^\dagger(2i\sqrt{n_0} \sin(\frac{gt}{2\sqrt{n_0}})) \hat{w}_{\beta_C(0)} \ket{\alpha} +  O\left(\frac{1}{n_0}\right)\\
      &=
\frac{1}{2\bar{n}+1}\exp\!\Big(-\frac{4n_0\sin^2\!\big(\frac{g t}{2\sqrt{n_0}}\big)}{2\bar{n}+1}\Big)+  O\left(\frac{1}{n_0}\right)
   \label{eq:rhoCoverlap}
    \end{aligned}
\end{equation}
As long as $gt \gtrsim 1$, the overlap becomes of order $  O\left(\frac{1}{n_0}\right)$ or smaller, as can be seen by expanding the for $gt/\sqrt{n_0}\ll 1$: $\text{Tr}[\hat{\rho}_C^{++} \hat{\rho}_C^{--}] \simeq \frac{1}{2\bar{n}+1}\exp\!\Big(-\frac{g^2 t^2}{2\bar{n}+1}\Big)+O\left(\frac{1}{n_0}\right)$. 

The reduced state of the cavity is identical to a convex mixture of the quantum states denoted as $\sum p_i \hat{\rho}_i$, for some set of probabilities $p_i$. The von-Neumann entropy of the latter generally satisfies \cite{Nielsen10}: 

\begin{equation}
\sum_i p_i S(\hat{\rho}_i) \leq S\left( \sum_i p_i \hat{\rho}_i \right) \leq \sum_i p_i S(\hat{\rho}_i) + H(p_i),
 \label{eq:VNentInq}
\end{equation}
 where $S(\rho_i)$ is the von Neumann entropy of state $\rho_i$, and $H(p_i)$ is Shannon entropy for the set of probabilities $p_i$. The upper bound of this inequality is saturated when the states $\rho_i$ have support on orthogonal subspaces \cite{Nielsen10}. In our case, the states $\hat{\rho}_i$ are $\hat{\rho}_C^{++}$ and $\hat{\rho}_C^{--}$, which have finite overlap amplitude during the measurement phase. The overlap $\text{Tr}[\hat{\rho}_C^{++} \hat{\rho}_C^{--}]$ vanishes for $gt\gtrsim 1$ as the states become completely orthogonal. This forces the entropy of the cavity to be given by the upper bound of Eq.~\eqref{eq:VNentInq}, up to terms $  O\left(\frac{1}{n_0}\right)$.
 Using that $S_C(0) = \log(\frac{(1+\Bar{n})^{1+\Bar{n}}}{\Bar{n}^{\Bar{n}}})$, we deduce:
 \bb \label{eq:SCtc}
 S_C(t_c) = \log(2\frac{(1+\Bar{n})^{1+\Bar{n}}}{\Bar{n}^{\Bar{n}}}) +   O\left(\frac{1}{n_0}\right),
 \ee
 and $S_C(t_c)-S_C(0)=\log 2 + O\left(\frac{1}{n_0}\right)$. 

  For $t_c\leq t \leq t_\text{min}$, the overlap $\text{Tr}\{\hat\rho_C^{++}\hat\rho_C^{--}\}$ remains negligible, such that $S_C$ still saturates the upper bound of Eq.~\eqref{eq:VNentInq}. Moreover, $\hat\rho_C^{++}(t)$ and $\hat\rho_C^{--}(t)$ evolve unitarily, such that $S_C(t)=S_C(t_c)$ and $S_C(t_\text{min})-S_C(t_c)\sim  O\left(\frac{1}{n_0}\right)$.\\
  
   While the expansion of the state $\hat\rho_\text{QC}$ can in principle be continued to evaluate the terms of order ${\cal O}(1/n_0)$, we find that the density operators $\hat\rho_C^{++}(t)$ and $\hat\rho_C^{--}(t)$ become non-Gaussian over times $t > 1/g$. Consequently, the approach followed in Appendix B for short times cannot be employed to describe the feedback regime.
 



\textbf{{\it Heat and work.}} We now exploit the definitions of heat and work introduced in the main text to analyze the thermodynamic behaviour of the cavity. We introduce the entropy $S[\hat{w}_{\beta(t)}]$ of the thermal state $\hat{w}_{\beta(t)}$ which has the same von Neumann entropy as $\hat{\rho}_C(t)$. It is parametrized by the mean photon number $\Bar{n}(t) = 1/(e^{-\beta_C(t) \hbar \omega_0}-1)$ via:

\begin{equation}
    \begin{aligned}
      S[\hat{w}_{\beta_C(t)}] &= \log[\frac{(1 + \bar{n}(t))^{1+\Bar{n}(t)}}{\bar{n}(t)^{\bar{n}(t)}}].
      \label{eq:meancav1}
    \end{aligned}
\end{equation}

Using Eq.~\eqref{eq:SCtc} and Eq.~\eqref{eq:meancav1}, we find
\begin{equation}
    \begin{aligned}
        \bar{n}(t) = 
\begin{cases}
    \bar{n}=\bar{n}(0) \leq \bar{n}(t) \leq \bar{n}(t_c)  & 0\leq gt < 1 \\
  \bar{n}(t_c)    &  1 \lesssim gt \lesssim \sqrt{n_0} \pi ,
   \label{eq:nbartC}
\end{cases}
    \end{aligned}
\end{equation}

where $\bar n(t_c)$ is the solution of:

\begin{equation}
    \begin{aligned}
      \log[\frac{(1 + \bar{n}(t_c))^{1+\Bar{n}(t_c)}}{\bar{n}(t_c)^{\bar{n}(t_c)}}] &= \log(2\frac{(1+\Bar{n})^{1+\Bar{n}}}{\Bar{n}^{\Bar{n}}})+O\left(\frac{1}{n_0}\right)
      \label{eq:nbartcexact}
    \end{aligned}
\end{equation}

Note that Eq.~\eqref{eq:nbartcexact} implies:
\begin{equation}
\begin{aligned}
  \bar{n} \leq \bar{n}(t_c) \leq 2\bar{n}+\frac{1}{2}.
  \label{eq:nbartcbounds}
\end{aligned}
\end{equation}

The heat exchanged by the cavity during interval $0$ to $t$ is

\begin{equation}
    \begin{aligned}
    Q_C(t) &= \text{Tr}[\hat{H}_C (\hat{\rho}_C(0) - \hat{\rho}(t))]\\
           &= \hbar \omega_0 (\bar{n} - \bar{n}(t))
           \label{eq:heatC0}
    \end{aligned}
\end{equation}

Using Eq.~\eqref{eq:nbartC} and Eq.~\eqref{eq:heatC0}:
\bb 
Q_C(t_c)\hbar \omega_0 (\bar{n} - \bar{n}(t_c)) \leq 0,
\ee
and $Q_C(t) = Q_C(t_c)$ for $t_c\leq t\leq t_\text{min}$.



When the qubit is initialized in the center of the Bloch sphere, no Rabi oscillation occur and the qubit energy remains constant for $t\in[0,t_\text{min}$. Moreover, the resonant Jaynes-Cummings interaction conserves the quantity $E_C+E_Q$, so that $E_C(t)$ is also constant over $[0,t_\text{min}$. From $\Delta E_C = -W_C-Q_C$, we find that the cavity performs an amount of work verifying:
\begin{equation}
    \begin{aligned}
        W_C(t)= -Q_C(t).
        \end{aligned}
\end{equation}


\textbf{{\it Von Neumann entropy of the qubit.}} The reduced state of the qubit is

\begin{equation}
    \begin{aligned}
        \hat{\rho}_Q(t) =
     \frac{1}{2} (\ket{+_y(t)}\bra{+_y(t)} + \ket{-_y(t)}\bra{-_y(t)}) + O(\frac{1}{n_0})
     \label{eq:rhoQM0}
    \end{aligned} 
\end{equation}

From Eq.~\eqref{eq:rhoQtc1}, we can see that during the time interval $[0,t_c]$, $\ket{\pm_y(t)} \approx \ket{\pm}$. Then , the expressions for $\ket{\pm_y(t)}$ reveals that during the feedback both $\ket{+_y}$ and $\ket{-_y}$ reaches $\ket{+_x}$. As a consequence of this, the overlap between $\ket{\pm_y(t)}$ remains negligible for $t\in[0,t_c]$ and increases up to $1- O(\frac{1}{n_0})$ at $t_\text{min}$. 



Upon interpreting Eq.~\eqref{eq:rhoQM0} as a convex sum of states over a given set of probabilities, we get at zero order in $1/n_0$:

\begin{equation}
    \begin{aligned}
        S[\hat{\rho}_Q(t)] = 
\begin{cases}
    \ln(2)  & 0 \leq gt < 1 \\
  0 \le S[\hat{\rho}_Q(t)] \le \ln(2)   & 1 \lesssim gt \lesssim \sqrt{n_0} \pi  
  \label{eq:entropyqubitMS}
\end{cases}
    \end{aligned}
\end{equation}

Finally, the quantum mutual information is obtained by summing the entropies of the qubit and the cavity, yielding at zero order in $1/n_0$:
\begin{equation}
    \begin{aligned}
      I_{QC}(0)&=  0\nonumber\\
      I_{QC}(t_c)&=  \log 2\nonumber\\
      I_{QC}(t_\text{min})&=  0.
    \end{aligned}
\end{equation}



\end{document}